\documentclass[a4paper,fleqn,usenatbib]{mnras}

\usepackage{newtxtext,newtxmath}

\usepackage[T1]{fontenc}
\usepackage{ae,aecompl}

\usepackage{graphicx}	
\usepackage{amsmath}	
\usepackage{amssymb}	
\graphicspath{{/Users/MattFong/Documents/LaTeX/Papers/Fong2018/Figures/}}

\def\ref@jnl#1{{\jnl@style#1}}
\def\aj{\ref@jnl{AJ}}				
\def\apj{\ref@jnl{ApJ}}			
\def\mnras{\ref@jnl{MNRAS}}		
\def\aap{\ref@jnl{A\&A}}			
\def\nat{\ref@jnl{Nature}}			
\def\prd{\ref@jnl{Phys.~Rev.~D}}	
\def\apjs{\ref@jnl{ApJS}}			
\def\pasj{\ref@jnl{PASJ}}			
\title[Weak Lensing: The Structure of Clusters on Large Scales]{Prospects for Determining the Mass Distributions of Galaxy Clusters on Large Scales Using Weak Gravitational Lensing}

\author[M. Fong et al.]{
M. Fong,$^{1}$\thanks{E-mail: mxf140930@utdallas.edu (MF)}
R. Bowyer,$^{4}$
A. Whitehead,$^{1}$
B. Lee,$^{1}$
L. King,$^{1}$
D. Applegate,$^{2}$
I. McCarthy$^{3}$
\\
$^{1}$Physics Department, The University of Texas at Dallas, 800 W. Campbell Road, Richardson, TX 75080, USA\\
$^{2}$Kavli Institute for Cosmological Physics, University of Chicago, 5640 South Ellis Avenue, Chicago, IL 60637-1433, USA\\
$^{3}$Astrophysics Research Institute, Liverpool John Moores University, 146 Brownlow Hill, Liverpool L3 5RF\\
$^{4}$Physics Department, Rice University, 6100 Main Street, Houston, TX 77005 USA
}

\date{Accepted XXX. Received YYY; in original form ZZZ}

\pubyear{2015}

\begin{document}
\label{firstpage}
\pagerange{\pageref{firstpage}--\pageref{lastpage}}
\maketitle

\begin{abstract}
For more than two decades, the Navarro, Frenk, and White (NFW) model has stood the test of time; it has been used to describe the distribution of mass in galaxy clusters out to their outskirts. Stacked weak lensing measurements of clusters are now revealing the distribution of mass out to and beyond their virial radii, where the NFW model is no longer applicable. 
In this study we assess how well the parameterised Diemer \& Kravstov (DK) density profile describes the characteristic mass distribution of galaxy clusters extracted from cosmological simulations.
This is determined from stacked synthetic lensing measurements of the 50 most massive clusters extracted from the Cosmo-OWLS simulations, using the Dark Matter Only run and also the run that most closely matches observations.
The characteristics of the data reflect the Weighing the Giants survey and data from the future Large Synoptic Survey Telescope (LSST).
In comparison with the NFW model, the DK model favored by the stacked data, in particular for the future LSST data, where the number density of background galaxies is higher.
The DK profile depends on the accretion history of clusters which is specified in the current study. Eventually however subsamples of galaxy clusters with qualities indicative of disparate accretion histories could be studied.

\end{abstract}


\begin{keywords}
gravitational lensing: weak -- galaxies: clusters: general -- dark matter 
\end{keywords}



\section{Introduction}
Galaxy clusters are the most massive bound structures in the Universe, providing key tests of 
our paradigm for structure formation and the cosmological model, as well as being powerful laboratories for investigating dark matter and the physics of baryons on large scales.

Most of the mass of clusters is thought to be cold dark matter, about 1/8th of the mass is X-ray emitting plasma, and only one or two percent is in the form of stars in cluster galaxies \citep*{Allen2011}. Combining multi-wavelength observations of the luminous components in clusters with gravitational lensing to access the total gravitating mass is essential to provide a full picture of these systems. 

The estimation of cluster mass underpins the use of clusters as cosmological probes. For example the mass profiles of clusters are sensitive to the properties of dark matter, testing whether it is cold and the interaction cross section (e.g. \citet*{Spergel2000}). The number of clusters as a function of mass and redshift - the cluster mass function - depends on the amount and nature of dark energy. As an extreme example, the existence of clusters at $z>1$ gives strong support for a significant energy density in dark energy e.g. \cite{Allen2011}. 

The analysis of the weak and strong gravitational lensing signatures of galaxy clusters is vital to the estimation of cluster mass density profiles and masses. Not all galaxy clusters have sufficient background multiply imaged sources or giant arcs to directly provide useful constraints on their central regions. Weak lensing operates over a much larger area of the sky. In the weak lensing regime, mass measurements can be carried out by fitting parametric models to the complex ellipticities describing the background weakly lensed galaxy shapes and orientations. Non-parametric mass reconstructions allow the mass to be mapped, particularly useful in the case of complex cluster systems such as major mergers and high redshift proto-clusters, both poorly described by parametric models. However, in general these non-parametric reconstructions are less well suited to cluster mass estimation because the shear field that is estimated from the background galaxy ellipticities has to be significantly smoothed, and the error properties of such maps are difficult to quantify (see \citet*{Hoekstra2013} for an overview of this topic).

\citet*{NFW1997} (Hereafter referred to as NFW) discovered that haloes formed in Cold Dark Matter (CDM) simulations of structure formation are well described by a universal form of density profiles that is characterised by a scale radius and the concentration of mass (or equivalent parameters). \citet*{Jing2000} found that about 70\% of haloes formed in CDM simulations are well fit by this so-called NFW profile. The NFW profile is also found to be a good fit to cluster scale haloes in more recent analysis of gravitational lensing data from the fields of real galaxy clusters, away from the central regions that are dominated by brightest cluster galaxies (e.g. \cite{Umetsu2016}). Variants on the NFW density profile that have slightly different behaviours, in particular in the inner and outer regions of clusters have been found to better reproduce high resolution haloes (e.g. \citet*{Moore1999}, \citet*{JingSuto2000}, \citet*{Fukushige2001}, Navarro et al. (2010)).

In general, although clusters are relatively simple and well described by the NFW model (with the caveat of setting aside complex mergers and cluster systems), nevertheless in their outskirts the large scale structure of filaments in the cosmic web and contributions from neighbouring clusters eventually become important. It has therefore long been recognised from simulations that eventually the NFW model ceases to be an accurate representation of the mass density. Formally the mass of an NFW halo diverges when integrated to infinity, and usually the NFW description is only applied out to the virial radius.
The halo model formalism (see e.g. \citet*{Cooray2002} for a review) is often used as a prescription for these additional contributions beyond the single halo.
Various works have used higher resolution simulations than were available at the time of \cite{NFW1997} and modified the form of the NFW profile in the inner regions of haloes and/or added an extra contribution to the density in the outskirts of haloes (e.g. \cite{Navarro2004}, \cite{Prada2006}, \cite{Hayashi2008}).
 Recently, using fits to cosmological simulations, \citet*{DK2014} (hereafter referred to as DK) suggested a new parametric model that describes clusters and the structure in which they are embedded, with a dependency on the rate at which mass has been accreted. 
 The splashback radius is a physically motivated definition for the extent of the halo, corresponding to the apocenter of the first orbits of particles after they have fallen into the halo. 
 Inside this radius particles are virialized, while outside the particles are infalling.
 This will naturally lead to a steepening of the density, where the splashback radius is where the density steepens most, which is related to the the mass accretion rate (\cite{More2015}, \cite{Umetsu2017}, \cite{Chang2017}, \cite{Snaith2017}). 
 We return to a full description of this model in the next section. The advent of detailed wide field images of galaxy clusters, and surveys that will cover many thousands of square degrees of the sky, motivates a consideration of the accuracy with which mass 
can be determined in the periphery of clusters, beyond the 1-halo term. \citet{Umetsu2017}, adopting the DK model, described lensing constraints on the shape of the average mass profile of 16 massive galaxy clusters from the CLASH sample of \citet*{Postman2012}. In that work they found that although the DK model was slightly preferred compared with the NFW that the differences were not statistically significant.

Looking forward to weak lensing data from future surveys such as LSST \citep*{Ivezic2008}, our primary goal in this paper is to establish if the DK profile will be distinguished from NFW using the average mass profile in the periphery of clusters, specifically detecting the density steepening at the outskirts of clusters. To this end we perform a synthetic stacking analysis of lensing observations of massive clusters extracted from the cosmo-OWLS simulations of \citet*{LeBrun2014}. We also consider synthetic data that reflects the characteristics of Weighing The Giants (hereafter WtG) (e.g. \citet*{Linden2014}).
In addition we briefly compare mass estimates for clusters derived by fitting NFW and DK models to azimuthally averaged weak lensing shear data. 

Throughout this work we use \textsc{Colossus} for all DK profile calculations \citep{DK2017} and \textsc{LmFit}  for all parameter estimates \citep{Newville2014}.

We discuss the simulation this method is applied to, and how we produce mock data that reflects what we may expect from surveys; in this paper we specifically used values to reflect WtG-like and LSST-like surveys. 
In Section \ref{sec:densityProfiles} and Section \ref{sec:lensing} we examine the NFW and DK profiles and their lensing properties. In Section \ref{sec:sims} we discuss the $WMAP7$ cosmology that we use throughout this paper. We then discuss the forms of the profiles in Section \ref{sec:forms}. We adopt the stacking methods from \cite{Niikura2015} to use weak lensing to determine the shape of weak lensing signals on average. 
In Section \ref{sec:fits} we give some details on mass estimates and the differences in the NFW and DK fits, and briefly discuss the impact of triaxiality in \ref{sec:triaxiality}. Further we show results of stacking with NFW scaling.

\section{Method: Stacked weak lensing shear signals of galaxy clusters}
\label{sec:Method}

\subsection{NFW and DK Density Profiles}
\label{sec:densityProfiles}
The NFW profile is a good fit to the spherically averaged profiles of haloes formed in cold dark matter simulations out to (very roughly the virial radius) $r_{200c}$ \citep{NFW1997}. 
It is therefore a commonly used mass density profile and is given by:
\begin{equation}
\rho_{\rm NFW}(r) = 
\frac{\delta_{c}\rho_{cr}}{\left(r/r_{s}\right)
\left(1+r/r_{s}\right)^{2}}, 
\label{eq:rhonfw}
\end{equation}
where the characteristic overdensity for the halo is given by 
\begin{equation}
\delta_{c}= \frac{200}{3}\frac{c^{3}}{\ln(1+c)-c/(1+c)}
\end{equation}
and the critical density of the universe is $\rho_{cr}(z) = \frac{3H^{2}(z)}{8\pi G}$. $H(z)$ is the Hubble parameter at redshift $z$, and $G$ is Newton's Gravitational constant. 
This profile is parameterised by the scale radius, $r_{s} = r_{200c}/c$, and the concentration parameter, $c$. 
$r_{200c}$ defines a sphere that encloses a mean density of $200\rho_{cr}(z)$, and the mass enclosed inside the sphere of radius $r_{200c}$ is
\begin{equation}
M_{200 c}  \equiv M(<r_{200c}) =  \frac{800\pi}{3} \rho_{cr}(z) r_{200c}^{3}.
\label{eq:M200c}
\end{equation}		
Another mass characteristic mass in this paper is
\begin{equation}
M_{200 m}  \equiv M(<r_{200m}) =  \frac{800\pi}{3} \rho_{m}(z) r_{200m}^{3},
\label{eq:M200m}
\end{equation}		
where $r_{200m}$ defines a sphere that encloses a mean density of $200\rho_{m}(z)$, and $\rho_{m}(z)$ is the mean matter density of the universe at redshift $z$.

DK proposes a mass density profile that describes the average density of clusters within and beyond the virial radius \citep{DK2014}. It more accurately captures the steepening at radii $r \geq 0.5r_{200m}$ of averaged $\Lambda$CDM haloes than the NFW profile, and flattens out to the mean density of the universe, $\rho_{\rm m}$ on large scales. 
The steepening is due to the splashback radius, $r_{sp}$, where accreted matter reaches the apocenter of their first orbit. This radius separates bound material and the infalling material.
Therefore $r_{sp}$ corresponds to when the slope of the density is steepest and is related to the mass accretion rate (\cite{More2015}, \cite{Umetsu2017}, \cite{Chang2017}, \cite{Snaith2017}). 
At this radius there is a sharp drop in density at the outskirts of halos, where $r_{sp}$ is the minimum of $dlog(\rho)/dlog(r)$. 
In \cite{Umetsu2017}, they place a lower limit on the splashback radius from their sample of 16 CLASH clusters, but the data do not identify a precise location. 

The DK density is 
\begin{align}
\begin{split}
\rho_{\rm dk}(r) = \rho_{\rm inner}*f_{trans} + \rho_{outer}
\\
\rho_{\rm inner} = \rho_{\rm Einasto} = \rho_{s}exp\left(-\frac{2}{\alpha}\left[\left(\frac{r}{r_s}\right)^{\alpha} - 1\right]\right)
\\
f_{\rm trans} = \left[1 + \left(\frac{r}{r_{t}}\right)^{\beta}\right]^{-\frac{\gamma}{\beta}}
\\
\rho_{\rm outer} = \rho_{\rm m}\left[ \frac{b_{\rm e}}{ \frac{1}{\rho_{max}}+\left(\frac{r}{5r_{\rm 200m}}\right)^{s_{\rm e}} }+1\right].
\end{split}
\label{eq:rhodk}
\end{align}
The Einasto profile \citet*{Einasto1965} describes the inner density, the transition term $f_{\rm trans}$ describes the steepening of the profile around a truncation radius $r_{\rm t}$, and the outer density is a power law that flattens out to the mean density of the universe. The inner density is characterised by the scale density $\rho_{\rm s}$, the scale radius $r_{\rm s}$, and $\alpha$, where $\rho_{\rm s}$ is the mass density at $r = r_{\rm s}$ and $\alpha$ determines how quickly the slope of the inner Einasto profile steepens. The transition term parameters are $\gamma$ and $\beta$, where $\gamma$ defines the steepness of the density around $r \approx r_{\rm 200m}$ and $\beta$ tells how quickly the slope changes. The outer density profile parameters are $\rho_{\rm m}$, the radius $r_{\rm 200m}$ that encloses an average overdensity of $200\rho_{\rm m}$, and parameters that describe the normalization and slope of the power law of the outer profile, $b_{\rm e}$ and $s_{\rm e}$ respectively. The outer density profile is a modification of \cite{DK2014}, where the term $\frac{1}{\rho_{max}}$ is introduced to avoid a spike toward the center of the cluster. This term determines the maximum overdensity that can be contributed from the outer profile. We use the \textsc{Colossus} \citep{DK2017} package for all DK profile calculations, where $\frac{1}{\rho_{max}} = 0.001$, and the input parameters for the DK profile are $M_{200c}$ and $c$. Note that the mean cosmic density acts like a constant density sheet of mass and hence does not impact on the shapes of distant galaxies.

 \citet{DK2014}  show that some of the parameters are correlated, reducing the number of free parameters from eight to four:
\begin{equation}
\begin{split}
\alpha(\nu) = 0.155 + 0.0095\nu^{2},
\\
r_{\rm t} = (1.9 - 0.18\nu)r_{\rm 200m},
\end{split}
\end{equation}
where $\nu$ is the peak height. In this paper we fix $\beta = 4$ and $\gamma = 8$, which is an accurate fit to the density profiles if the truncation radius is related to $\nu$ and $r_{200m}$ \citep{DK2014}. The remaining four parameters are the Einasto parameters $\rho_{\rm s}$ and $r_{\rm s}$, and the two outer profile parameters $b_{\rm e}$ and $s_{\rm e}$. In \cite{DK2014}, the best-fit for the outer parameters are $b_{\rm e} \approx 1.0$ and $s_{\rm e} \approx 1.5$, which we fix for this paper. Therefore, in Section \ref{sec:fits} we only fit for $M_{\rm 200c}$ and $c$, which are related to the inner density parameters $\rho_{\rm s}$, $r_{\rm s}$, and $r_{200m}$ \citep{Diemer2013}. We use the \textsc{Colossus} code for conversions between mass definitions \citep{DK2017}.


\subsection{Weak Gravitational Lensing}
\label{sec:lensing}

In this paper we study the average weak lensing signals of clusters of galaxies, and we use spherically symmetric density profiles to describe them. With a spherically symmetric 3D mass density profile, we obtain the 2D surface mass density by integrating along the line-of-sight, $dz$: 
\begin{equation}
\Sigma(R) = 2 \int_{0}^{\infty}\rho(R,z)dz.
\label{eq:Sigma}
\end{equation}
$R = D_d \sqrt{\theta_1^2 + \theta_2^2}$ is the projected radius relative to the center of the lens on the lens plane, where $\theta_{1}$ and $\theta_{2}$ are angular variables on the sky. 

The convergence is the ratio of the surface mass density to the critical surface mass density:
\begin{equation}
\kappa(R) = \frac{\Sigma(R)}{\Sigma_{cr}},
\label{eq:kappa}
\end{equation}
where the critical mass density is \citep*{Subramanian1986}:

\begin{equation}
\Sigma_{\rm cr} \equiv \frac{v_c^2}{4\pi G} \frac{D_s}{D_d D_{ds}}.
\label{sigmaCrit}
\end{equation}
$D_d$, $D_s$, and $D_{ds}$ are the angular diameter distances between the observer and the lens, the observer and the source, and the lens and the source respectively, while $v_c$ is the speed of light, and G is the gravitational constant.

In the case of a spherically symmetric lens, the shear is given by 
\begin{equation}
\gamma(x) = \frac{\overline{\Sigma}(x) - \Sigma(x)}{\Sigma_{ \rm cr}(z_\mathrm{d},z_\mathrm{s})}
\equiv  \frac{\Delta\Sigma(x)}{\Sigma_{\rm cr}(z_\mathrm{d},z_\mathrm{s})},
\label{eq:gamma}
\end{equation}
where $x = R/r_{\rm s}$, with $R$ being the projected distance on the lens plane from the halo centre. The mean surface mass density of the halo is given by 
\begin{equation}
\overline{\Sigma}(x)= \frac{2}{x^2} \int_0^x x' 
\Sigma(x') dx' .
\label{sigmaBar}
\end{equation}

The NFW tangential shear is 
\begin{equation}
\gamma^{\rm NFW}(x) 
\equiv  \frac{\Delta\Sigma^{\rm NFW}(x)}{\Sigma_{\rm cr}(z_\mathrm{d},z_\mathrm{s})}
= \frac{2\rho_cr_s}{\Sigma_{\rm cr}(z_\mathrm{d},z_\mathrm{s})}f^{\rm NFW}(x),
\label{eq:gammanfw}
\end{equation}
with the form of the shear (\cite{Niikura2015}, \cite{Bartelmann1996} )
\begin{equation}
f^{\rm NFW}(x)\nonumber \\
  =\left\{
  \begin{array}{ll}
   {\displaystyle \frac{2}{x^2}\ln\frac{x}{2}+\frac{1}{1-x^2}
    \left(
1+\frac{2-3x^2}{x^2\sqrt{1-x^2}}\mathrm{cosh}^{-1}\frac{1}{x}
	     \right)},
    & (x<1)\\
   {\displaystyle \frac{5}{3}-2\ln2}, &  (x=1)\\
      {\displaystyle \frac{2}{x^2}\ln\frac{x}{2}-\frac{1}{x^2-1}
    \left(
1+\frac{2-3x^2}{x^2\sqrt{x^2-1}}\mathrm{cos}^{-1}\frac{1}{x}
	     \right)},
 & (x>1)\\
  \end{array}
			 \right.
\label{eq:fnfw}
\end{equation}

and the central density  
\begin{equation}
\rho_{c} = 
\frac{M_{200c}}{4 \pi r_s^3 m(c)}, 
\label{eq:rhoc}
\end{equation}
where $m(c) = log(1+c) - \frac{c}{1+c}$.

\begin{figure}
	\includegraphics[width=\columnwidth]{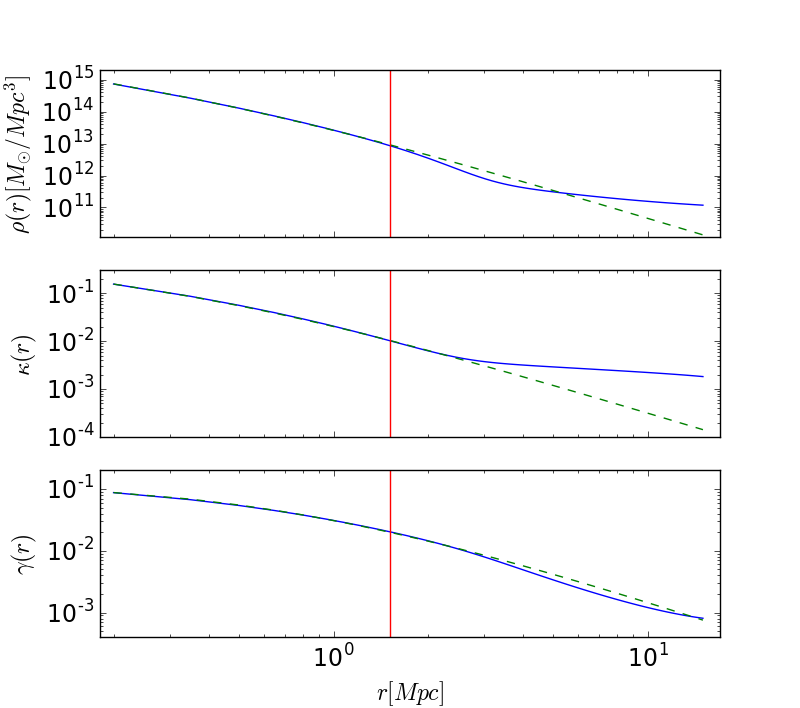}
	\caption{For NFW and DK, the upper, middle, and lower panels compare the 3D mass density, convergence, and shear profiles respectively. The DK profile is the solid blue curve, the NFW is the dashed green curve, and $r_{200c} = 1.41 Mpc$ is the solid red vertical line. In this example we use typical values of $M_{200c} = 5 \times 10^{14} M_\odot$ and $c = 4$ to illustrate that the DK and NFW profiles are a good match out to the virial radius.
	}
	\label{fig:rhoKapGam}
\end{figure}

\begin{figure}
	\includegraphics[width=\columnwidth]{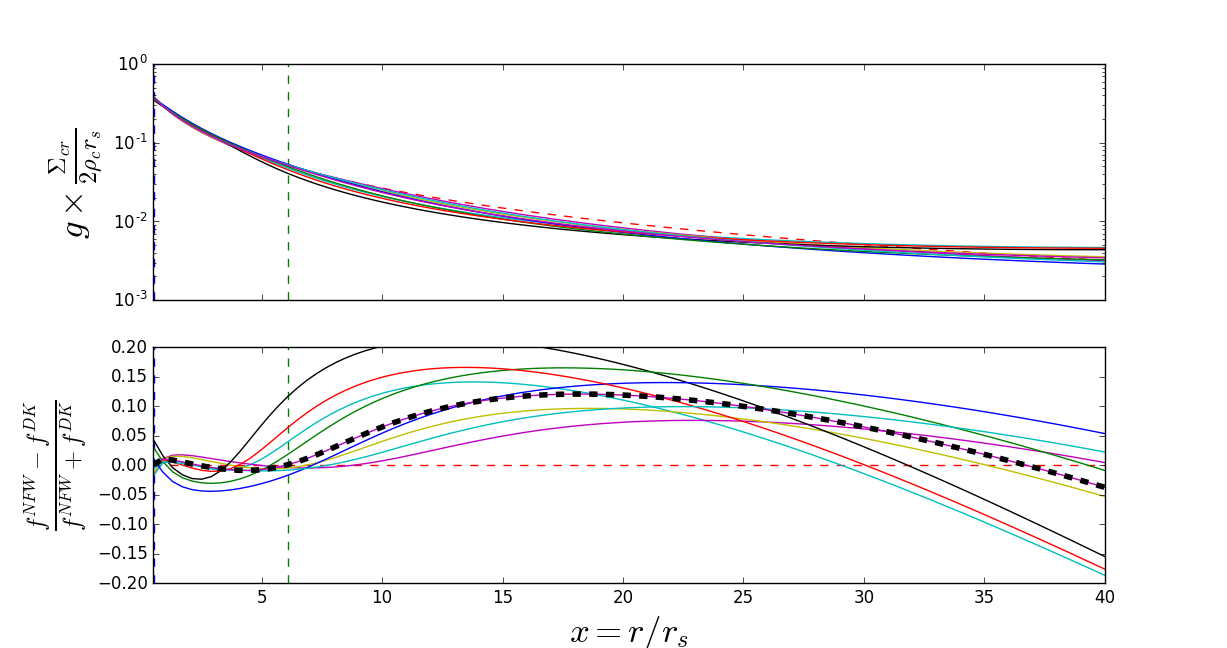}
	\caption{The top panel here shows the shear forms (scaled shear) for the NFW and DK profiles for a wide range of parameters. The dashed red curve is the NFW form, $f^{NFW}(x)$, the solid curves are the DK forms, $f^{DK}$, and the dashed vertical line a outer fit radius of $2.3$ Mpc ($r_{s} \approx 0.38$ Mpc). The range of parameters plotted are $2.0 < M_{200c} [10^{14} M_{\odot}] < 17.4$ and $2.5 < c < 5.5$. The NFW form is self-similar so for any parameters the form will lie on the dashed curve, however the DK form varies with mass and concentration.
	The bottom panel shows the fractional difference between the NFW and DK forms. The dashed horizontal line is the NFW form (y = 0), and the solid lines are the DK forms. Most of parameters used in these panels are quite extreme, where a more typical mass and concentration ($M_{200c} [10^{14} M_{\odot}] = 5.0$, $c = 4.0$) is highlighted by a thick black dashed curve. To the left of the dashed vertical line, the DK forms are nearly self-similar for more typical mass ranges used in this paper. 
	This shows that, though the DK and NFW forms differ quite significantly for large radii, for typical weak lensing fit ranges the DK and NFW forms nearly agree with one another for more typical masses and concentrations we use in this paper.}
	\label{fig:fdk_vs_fnfw}
\end{figure}

The DK shear is non-analytic, so it must be calculated numerically. For typical parameters that are relevant to this paper ($M_{200c} = 5 \times 10^{14} M_\odot$ and $c = 4$; $r_{\rm 200c} = 1.41$ Mpc), Figure \ref{fig:rhoKapGam} shows that the DK and NFW 3D density, convergence, and shear are very similar within $r_{\rm 200c}$. The profiles remain in agreement for a wide range of mass and concentration combinations within $r_{\rm 200c}$, encompassing the average parameters for the sample of 50 massive simulated clusters that are used in this work. 
We hypothesize that the DK shear can be approximated by:
\begin{equation}
\gamma^{\rm DK}(x) 
\equiv  \frac{\Delta\Sigma^{\rm DK}(x)}{\Sigma_{\rm cr}(z_\mathrm{d},z_\mathrm{s})}
= \frac{2\rho_cr_s}{\Sigma_{\rm cr}(z_\mathrm{d},z_\mathrm{s})}f^{\rm DK}(x, \vec{\pi}),
\label{eq:gammadk}
\end{equation}
where $\vec{\pi}$ are the mass and concentration parameters. Since the DK profile, $f^{DK}(x, \vec{\pi})$ has to be calculated numerically. 
In Figure \ref{fig:fdk_vs_fnfw} we vary the parameters for the NFW and DK shear forms. 
The range of parameters plotted are $2.0 < M_{200c} [10^{14} M_{\odot}] < 17.4$ and $2.5 < c < 5.5$, where the mass range is determined from the minimum and maximum true masses from the simulations, and the concentration range encompasses the minimum and maximum concentrations using the c-M relation of various models\footnote{\textsc{Colossus} was used to calculate concentrations. The models are Bullock et al. 2001, Duffy et al. 2008, Klypin et al. 2011 , Prada et al. 2012, Bhattacharya et al. 2013,  Dutton \& Macci\`{o} 2014, Diemer \& Kravtsov 2015, Klypin et al. 2016. Please see \citet{DK2017} for details.}.
In the top panel we show the NFW and DK shear forms, dashed curve and solid curves respectively, and a weak lensing outer fit radius of $2.3$ Mpc ($r_{s} \approx 0.38$ Mpc), dashed vertical line.
Because the NFW shear can be written exactly as in Equation \ref{eq:gammanfw}, the NFW forms are in exact agreement, and is represented by the dashed red curve $f^{NFW}(x)$. 
However since Equation \ref{eq:gammadk} is an approximation, the $f^{DK}(x, \vec{\pi})$ forms do not lie on top of each other for the wide range of parameters listed above. This means that $f^{DK}(x, \vec{\pi})$ in Equation \ref{eq:gammadk} depends on mass and concentration. 
A more typical mass and concentration in the simulation ($M_{200c} [10^{14} M_{\odot}] = 5.0$, $c = 4.0$) is represented by a thick dashed curve. 
The DK forms, for masses and concentrations we use in this paper, have roughly the same amplitude as the thick dashed black curve within the weak lensing fit range.
For cosmo-OWLS (Section \ref{sec:sims}) the mean mass of the 50 most massive clusters is $\langle M^{3D}_{200c} \rangle = 6.64 \times 10^{14} M_{\odot}$ with a range $4.04 <  M^{3D}_{200c} [10^{14} M_{\odot}] < 17.40$\footnote{The simulation masses ($M^{3D}_{200c}$) are calculated by identifying the particle deepest in the potential well of a cluster and calculating the density within spheres around that particle. When a mean enclosed density of $200\rho_{c}$ is reached, the mass enclosed in $r^{3D}_{200c}$ is $M^{3D}_{200c}$.}.
Therefore the plot shows that the DK and NFW forms are self-similar over fit ranges typically probed by weak lensing observations, for clusters relevant to this work. 
Furthermore, \cite{Niikura2015} shows that the NFW shear does a good job at representing the stacked shear of the 50 massive clusters in their simulations and in observations, out to about $2.3$ Mpc. 
With the NFW and DK shear forms in agreement within roughly $r_{200c}$, we can predict the mass and concentration of stacked shear by fitting over the weak lensing range (where both the DK and NFW shear profiles are nearly self-similar) using the NFW profile. With this we predict the DK form out to larger radii and compare that with stacked shear data within and beyond $r_{200c}$.

On the sky we observe the (complex) lensed ellipticities of background galaxies:
\begin{equation}
\epsilon = \frac{\epsilon^{s} + g}{1 + g^{*}\epsilon^{s}},
\label{eq:ellipticity}
\end{equation}
where $\epsilon^{\rm s}$ is the intrinsic ellipticity of the background galaxy, and the reduced shear is 
\begin{equation}
g = \frac{\gamma}{1 - \kappa}.
\end{equation}

In the absence of lensing, if galaxies are randomly oriented in the universe, taking the average ellipticities of enough unlensed galaxies on a patch of sky results in $\langle \epsilon^{\rm s} \rangle \approx 0$. In the weak lensing limit, the average of the observed lensed ellipticities then yields \citep*{Schneider2005}
\begin{equation}
\langle \epsilon \rangle = g.
\label{eq:avgEllipticities}
\end{equation}

For spherical profiles $g$ is given by
\begin{equation}
g(x) = \frac{2\rho_cr_s}{\Sigma_{\rm cr}(z_\mathrm{d},z_\mathrm{s})}
\frac{f(x)}{1 - \kappa(x)}
= \frac{2\rho_cr_s}{\Sigma_{\rm cr}(z_\mathrm{d},z_\mathrm{s})}F(x).
\label{eq:g}
\end{equation}
Here $f(x)$ can be the form of the NFW or DK shear profiles, $f^{\rm NFW}(x)$ and $f^{DK}(x, \vec{\pi})$ respectively, while $F(x)$ is the form of the NFW or DK reduced shear profiles, $F^{\rm NFW}(x)$ and $F^{DK}(x, \vec{\pi})$ respectively. In the weak lensing regime $\kappa(x) \ll 1$ and $F(x) \approx f(x)$.

\begin{figure}
	\includegraphics[width=\columnwidth]{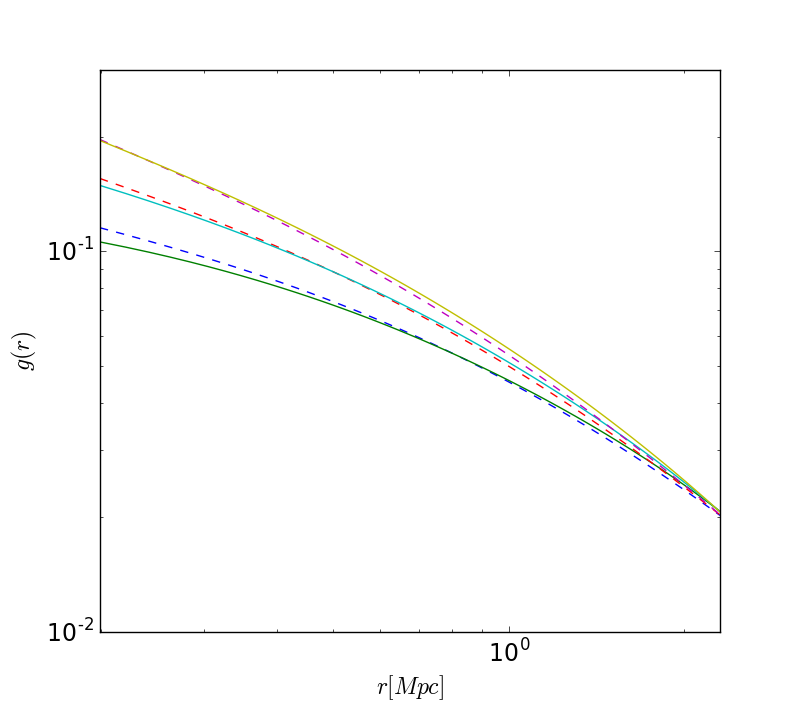}
	\caption{The solid and the dashed lines are the DK and NFW ideal reduced shear profiles respectively. Though the profiles are similar for the typical average parameters used in this paper, we show that for higher mass bins there are deviations when varying concentration. The mass of these profiles is $M_{200c} = 1\times 10^{15} M_{\odot}$ with concentrations $3$, $4$, and $5$, from bottom to top. Six of the 50 clusters used in this paper are more massive than this example. 
	}
	\label{fig:hiMassBin}
\end{figure}

Figure \ref{fig:hiMassBin} shows the ideal reduced shear profiles of the DK and NFW profiles with the same parameters. The solid lines are DK and the dashed are the NFW. The mass of these profiles is $M_{200c} = 10\times 10^{14} M_{\odot}$ with concentrations $3$, $4$, and $5$, from bottom to top. This shows that the DK ideal reduced shear profiles do in fact differ from the NFW for high masses, well outside of the mean masses of clusters in this paper. However, for more typical masses and concentrations, $M_{200c} \approx 5 - 7 \times 10^{14} M_{\odot}$ and $c = 2 - 7$, the inner profiles within $r_{200c}$ agree with one another.

\subsection{Simulations of Synthetic Lensing Catalogues}
\label{sec:sims}
In this subsection we describe the simulations from which cluster mass haloes are extracted. Cosmo-OWLS \citep{LeBrun2014} is a suite of cosmological hydrodynamic simulations using a version of the GADGET3 code, with runs exploring structure formation in cosmologies with dark matter only, and with dark matter and various prescriptions for baryonic physics. The simulation runs were carried out in periodic boxes of 400 $h^{-1}$ Mpc on a side (comoving), using the same initial conditions and cosmological parameters from $WMAP7$ \citep{wmap7}: $\{\Omega_m,\Omega_b,\Omega_{\Lambda},\sigma_8, n_s, h\} = \{0.272,0.0455,0.728,0.81,0.967,0.704\}$. Each run uses $2\times1024^3$ particles with masses $\approx 3.75 \times 10^9 h^{-1} \mathrm{M_{\odot}}$ and $\approx 7.54 \times 10^8 h^{-1} \mathrm{M_{\odot}}$ for dark matter particles and for baryonic particles respectively. 

In this paper we use two of the six simulation runs from \citet{LeBrun2014}: 
\begin{itemize}
	\item DMO: a dark matter-only run that accounts for only gravitational interaction between particles.
	\item AGN 8.0: in addition to gravity this run implements star formation \citep{Vecchia2008}, with radiative cooling, stellar evolution and chemical enrichment (\cite{Wiersma2009a}, \cite{Wiersma2009b}), supernova feedback \citep{Vecchia2008} and a UV/X-Ray photoionizing background \citep{Haardt2001}. The growth of supermassive black holes and AGN feedback are described by prescriptions from \citet{Booth2009}. AGN 8.0 yields clusters that are closest to observed clusters out of the various Cosmo-OWLS runs. 
\end{itemize}
The 50 most massive clusters at $z=0.25$, having M$_{200c}$ in excess of $4.04 \times 10^{14} \mathrm{M_{\odot}}$ in the DMO run, were extracted in boxes of 30 Mpc on a side (see Section ~\ref{sec:densityProfiles}). These clusters were matched to their counterpart clusters in the AGN 8.0 run, then each of the clusters was projected along the z-axis in order to obtain maps of the projected mass density, a scaled version of the lensing convergence. The maps of lensing shear and reduced shear were obtained using the relationship between lensing convergence and shear in Fourier space(see e.g. \cite{Schneider2005}).

Sets of background source galaxies with random locations were generated, assuming $z=1$. Different number densities of galaxies were explored, $n_{0} = 10$ or $30$ gal/arcmin$^2$  and the total number of galaxies adjusted using Poisson statistics to allow for variations. For upcoming surveys like LSST, we use $n_{0} = 30$ gal/arcmin$^2$, while for past surveys like WtG, $n_{0} = 10$ gal/arcmin$^2$ in keeping with the observations. The (complex) ellipticity of each source galaxy, describing its shape and orientation, $\epsilon$ (Equation ~\ref{eq:ellipticity}), where the intrinsic ellipticity $\epsilon^{s}$ was drawn from a Gaussian distribution with $\sigma_{\epsilon}=0.25$. 
Including only an ellipticity dispersion is optimistic and our synthetic measurements of lensing do not include contributions from galaxy shape measurement errors. 
For observations, the effective shear dispersion can be as large as $0.4$.
Following \citet{Schneider2000}, the number density of galaxies on each small patch of sky was modified to account for lensing magnification\footnote{Although this effect is small, we include it for completeness.}, $n = n_{0}\mu^{\beta-1}$, where $\mu$ is the magnification on the small patch, and $\beta$ is the slope of the unlensed source counts taken to be  $\beta=0.5$. The magnification of each galaxy due to the foreground cluster was calculated at each location, and after drawing a random uniform variate $\eta \in [0,1]$, the galaxy was only included in the lensed galaxy catalogue if $[\mu]^{\beta - 1} \geq \eta$, otherwise discarded. This accounts for the change in brightness of lensed galaxies as well as the stretching of space around them.

From the simulation we obtain $\kappa$, $\gamma$ (real and complex components $\gamma_1$, $\gamma_2$) at each pixel location, then taking into account survey characteristics, obtain (complex) $\epsilon$ (including shape noise) at galaxy positions on the lens plane. 
The ellipticity amplitude and tangential component, $\epsilon_{\rm (s_{\rm a})}$ and $\epsilon_{\rm (s_{\rm a})+}$ respectively, are given by \citep{Schneider2000}
\begin{equation}
\epsilon_{\rm (s_{\rm a})} = 
\sqrt{ \epsilon_{\rm 1(s_{\rm a})}^2 + \epsilon_{\rm 2(s_{\rm a})}^2}
\label{eq:ellipticityAmplitude}
\end{equation}
and 
\begin{equation}
\epsilon_{\rm (s_{\rm a})+} = 
-\epsilon_{\rm 1(s_{\rm a})}cos(2\phi) - \epsilon_{\rm 2(s_{\rm a})} sin(2\phi),
\label{eq:tangentialEllipticity}
\end{equation}
where $ \epsilon_{\rm 1(s_{\rm a})}$ and $ \epsilon_{\rm 2(s_{\rm a})}$ are the real and complex components of the lensed galaxy $s_{a}$ respectively, while $\phi$ is the angular position of the lensed galaxy on the lens plane. 

For the AGN 8.0 and DMO simulations multiple WtG and LSST-like runs were carried out, where the differences between similar survey-like runs are the random background source galaxy locations (at $z=1$) and shape noise realizations. 
This was done to see if different realizations would give significantly different results, or stacked cluster shape. Though all the LSST-like runs had some large differences in mass estimations for individual clusters, there were no significant differences in the statistical results. The same can be said about the WtG-like runs, but their results have larger error bars (see Sec.~\ref{sec:WtG}). We also explore using exclusively ideal DK or NFW haloes for each set of survey-like parameters to compare our simulation runs with ideal runs (see Section ~\ref{sec:idealDKHalos})


\subsection{NFW and DK reduced shear forms}
 \label{sec:forms}
In this paper we want to see if the numerical DK form $F^{DK}(x, \vec{\pi})$ is supported by surveys on large scales, keeping in mind the concentration dependency found in Section \ref{sec:lensing} outside of $r_{200c}$. In that section we showed that, within the weak lensing regime and $r_{200c}$, the NFW and DK profiles behave similarly. So if the concentration can be determined by using the NFW profile, then that concentration can be used in the DK form.
As a reference we will compare this to the NFW form $F^{\rm NFW}(x)$. For the NFW, the reduced shear $g^{\rm NFW}(x) = \frac{\gamma^{\rm NFW}(x)}{1 - \kappa^{\rm NFW}(x)}$ is, though complicated, an analytic function of $x$. Note that $g(x)$ and $x$ are dimensionless. 

From Eq.~\ref{eq:g}, with an ideal spherical halo we can get the "form" of the reduced shear for an individual halo with 
\begin{equation}
F(x) = \frac{\Sigma_{\rm cr}}{2\rho_c r_s}g(x).
\label{eq:F}
\end{equation}
The scaling factor $\frac{\Sigma_{\rm cr}}{2\rho_c r_s}$ scales the signal according to cluster mass. So in the ideal case, the form of the reduced shear, $F(x)$, will be the same for all clusters in the weak lensing regime and $r_{200c}$ and within the parameter space of this paper. The form, $F(x) = \frac{f(x)}{1 - \kappa(x)}$, can vary due to the parameter dependency in $\kappa(x)$. However, our focus is in the weak lensing regime where $\kappa(x) \ll 1$ and therefore $F(x) \approx f(x)$\footnote{It can be shown that scaling $\gamma^{NFW}$ for various ideal haloes will all conform to the curve $f^{NFW}(x)$ for all of $x$}. For the rest of this paper we will simply just refer to $F(x)$ since $F^{DK}(x, \vec{\pi})$  and $F^{NFW}(x)$ are very similar in the weak lensing regime out to the virial radius. We want to test if the DK profile can describe stacked clusters beyond the virial radius.
As the scaling factor shifts a reduced shear signal vertically to obtain $F(x)$, the choice in scaling the radial bins can shift the curve either left or right. It is important to note that neither changes the shape of the signal. Through this paper we scale the radial bins by $r_{s}$.

\subsection{Fitting Method}
\label{sec:fitMethod}
To obtain the parameters that describe the profiles in Section \ref{sec:densityProfiles}, we use the background galaxy ellipticities, described in Section \ref{sec:lensing}. This is done by azimuthally averaging the tangential shear with $N = 300$ galaxies per bin to obtain $g(r)$ (Eq.~\ref{eq:avgEllipticities}) for cluster $a$, then fitting the data with $g^{NFW}(r)$ to obtain the free parameters $(M_{200c}, c)$ for each cluster, or $(M_{(a)}, c_{(a)})$. Throughout this paper we fit with the NFW profile since the fitting process is much faster than using the DK profile (NFW shear is analytic) and the parameter estimation is similar to the DK profile anyway.
With these parameters we can calculate the scaling factor $\frac{2\rho_c r_s}{\Sigma_{\rm cr}}$ in Eq.~\ref{eq:F} using Eq.~\ref{eq:M200c}, Eq.~\ref{eq:rhoc} and $r_{s} = r_{200c}/c$. The fit ranges are set to $0.20 < r [$Mpc$] < 2.30$ and $0.75 < r [$Mpc$] < 3.00$ for LSST-like surveys and WtG-like surveys respectively. The LSST-like lower limit of the fit range was taken from \cite{Niikura2015} while the higher fit range is from the highest value of $r^{3D}_{200c}$ from the simulation catalog. The fit range from the WtG sample is from \cite{Applegate2014}. 
The two fit ranges used in this paper do not have a significant impact on the stacking results.
The error for the fits is determined from the variance from binning and the number of galaxies per bin, $\sigma / \sqrt{N}$. Throughout this paper we use the public module \textsc{LmFit}\footnote{The \textsc{LmFit} package is Free software, using an Open Source license} as the fitting tool \citep*{Newville2014}.
The fitting function used in LmFit minimizes the sum of squared residuals:
\begin{equation}
\mathcal{L} = \sum_{i} \frac{ (\langle \epsilon \rangle_{i} - g^{NFW}_{i})^{2}}{\sigma_{i} / \sqrt{N_{i}}},
\end{equation}
where $\langle \epsilon \rangle_{i}$, $g^{NFW}_{i}$, and $\sigma_{i} / \sqrt{N_{i}}$ are the average tangential ellipticities, predicted NFW reduced shear, and error in bin {i}


\subsection{Stacking Cluster Signals Without Scaling}
\label{sec:stackingWithoutScaling}
Realistic clusters are not spherical, and mass estimates from weak lensing fits using a spherical model generally have high scatter with small bias. However, we can determine the mean signal of clusters better than any individual cluster which can have low signal to noise for individual measurements, by taking the weighted average of many clusters' signals - a method called stacking. This process increases the signal to noise by averaging out any shear due to substructure unrelated to the lens, and the impact of triaxiality. Though we lose information of individual clusters with this method, we can obtain more precise estimates on the average behaviour of the stacked clusters. In this paper we follow the method in \cite{Niikura2015}: 
\begin{equation}
\langle \Delta \Sigma (R) \rangle=
\frac{1}{N}\sum_{a=1}^{N_c}\sum_{s_a; |{\bf R}_{(a)s_a}|\in R'}
w_{(a, s_a)} \Sigma_{\rm cr(a)}
\epsilon_{(s_a)+}({\bf R}_{s_a})
\label{eq:stackNiikura}
\end{equation}
where the first summation $\sum_{a=1}^{N_{\rm c} }$ is over each cluster $a$ in the stack with $N_{c}$ clusters, while the second $\sum_{s_a; |{\bf R}_{(a)s_a}|\in \rm R'}$ runs over the background galaxies $s_{\rm a}$, that belong to cluster $a$, that reside in the preset radial bins $R'$. The tangential ellipticity of the $s_{\rm a}$th source galaxy of cluster $a$ at position ${\bf R}_{s_a}$ is $e_{(s_a)+}({\bf R}_{s_a})$. The normalization factor is defined as 
\begin{equation}
N = \sum_{a=1}^{N_{c} }\sum_{s_a; |{\bf R}_{(a)s_a}|\in R'} 
w_{(a, s_{a})},
\label{eq:normalizationFactor}
\end{equation}
with the weight factor for each background galaxy in a cluster adopted from \cite{Okabe2016}\footnote{This differs slightly from \cite{Niikura2015} by the ellipticity amplitude $\epsilon_{(s_a)}^2$. The ellipticity amplitude in the weight is not included in this paper.}:
\begin{equation}
w_{(a,s_a)}=\frac{1}
{\Sigma_{\rm cr}^2(z_a,z_{s_a})(\sigma_{(s_a)\epsilon}^2+\alpha^2)}. 
\label{eq:weight}
\end{equation} 
$z_a$ and $z_{s_a}$ are the redshifts of the $a$th cluster and $s_a$th background galaxy respectively. For the $s_a$th source galaxy, $\sigma_{(s_a)\epsilon}$ is the measurement error, and $\alpha$ is the constant factor that regularizes the weight. 
For this paper we set $\sigma_{(s_a)\epsilon}$ to a constant for all galaxies and $\alpha = 0.4$ \citep[Similar to ][]{Niikura2015, Okabe2010a} in Eq.~\ref{eq:weight}. With $\Sigma_{\rm cr(a)}$ a constant throughout this paper (redshifts are constant), this makes the weight a constant value and will be factored out with $N$. In practice the redshifts and measurement uncertainties, $\sigma_{(s_a)\epsilon}$, would be provided from observational data. 

The stacked radial bins is given as
\begin{equation}
R \equiv \frac{1}{N}\sum_{a=1}^{N_c}\sum_{s_a; |{\bf R}_{(a)s_a}|\in R'}
w_{(a, s_a)}R_{(a)s_a},
\label{eq:stackR}
\end{equation}
where $R_{(a)s_a}$ is the position of the background galaxy $s_a$ from the center of cluster $a$.\footnote{In this paper we use a common redshift of clusters and of sources, so $\Sigma_{\rm cr(a)}$ essentially becomes factored out. So we are taking the weighted average positions of sources in the bins set by $R'$.}

The statistical uncertainty of the stacked lensing at each radial bin is estimated as
\begin{equation}
\sigma_{\langle \epsilon \rangle}^2(R)=\frac{1}{2N^2}
\sum_{a=1}^{N_c}\sum_{s_a; |{\bf R}_{(a)s_a}|\in R'} 
w_{(a,s_a)}^2 \epsilon_{(s_a)}^2({\bf R}_{s_a}).
\label{eq:sigmaEllipticities}
\end{equation}

It is important to note that Equations \ref{eq:stackNiikura} and \ref{eq:sigmaEllipticities} are functions of stacked radial bins $R$ (Eq.~\ref{eq:stackR}).

 
\subsection{Stacking Cluster Signals With Scaling}
\label{sec:stackingWithScaling}

\cite{Niikura2015} shows that stacking with NFW scaling will have less scatter as opposed to the stacking without scaling method and that the NFW profile describes stacked clusters (with or without scaling) out to the virial radius very well. Furthermore, for reasonable parameters in stacking, the DK and NFW profiles ($\rho$, $\Sigma$, $\gamma$, and $g$) agree with each other out to the virial radius. With this process, we can scale the lensing data with the common scaling factor $\frac{\Sigma_{\rm cr}}{2\rho_c r_s}$ (in Eq.~\ref{eq:F}) and study the non-analytic DK reduced shear form $F^{DK}(x, \vec{\pi})$. The parameters $\vec{\pi}$ (in the form, see Section \ref{sec:lensing}) is determined by fitting the stack, Equation \ref{eq:stackNiikura}, with the NFW profile.

Stacked weak lensing with NFW scaling reduces scatter of the reduced shear signals going into the stack. 
Haloes in simulations exhibit a high degree of self-similarity (e.g., NFW) when scaled appropriately. Therefore the expectation when we scale before stacking is that the diversity (or spread) in the profiles will be minimised \citep[][Fig.4]{Niikura2015}. 

The stack with NFW scaling will be represented as $\langle F(x) \rangle$, instead of the individual halo form $F(x)$ (Eq.~\ref{eq:F}). The stacked reduced shear with scaling follows as
\begin{equation}
\langle F(x) \rangle = 
\frac{1}{N}\sum_{a=1}^{N_{\rm c}}\sum_{s_{a};| {\bf x}_{(a)s_a}|\in x'}
\frac{w_{\rm (a, s_{\rm a}) } \Sigma_{\rm cr(a)} \epsilon_{\rm (s_{\rm a})+}(\textbf{x}_{\rm s_{\rm a} }) }
{2\rho_{\rm c} ( M_{\rm (a)}, c_{\rm (a)}) r_{\rm s}( M_{\rm (a)}, c_{\rm (a)})},
\label{eq:stackForm}
\end{equation}

where $\epsilon_{\rm (s_{\rm a})+}(\textbf{x}_{\rm s_{\rm a} })$ is the tangential ellipticity of a source galaxy at position $\textbf{x}_{\rm s_{\rm a}} = R/r_{\rm s}( M_{\rm (a)}, c_{\rm (a)})$, and the parameters $(M_{\rm (a)}, c_{\rm (a)})$ are the NFW $M_{200c}$ and concentration fit parameters for cluster $a$ (see Sec.~\ref{sec:fits} for details on fits). In the case of stacking with scaling, the second summation in the normalization $N$ is over ${s_{a};| {\bf x}_{(a)s_a}|\in x'}$.

\begin{equation}
x \equiv \frac{1}{N}\sum_{a=1}^{N_c}\sum_{s_a; |{\bf x}_{(a)s_a}|\in x'}
w_{(a, s_a)}x_{(a)s_a}, 
\label{eq:stackx}
\end{equation} 
\cite{DK2014} (Figure 3) shows that scaling the density profiles by $\rho_{m}$ and the radial bins by $r_{200m}$ can reduce scatter for the outskirts of haloes. We studied the stacking results with $r_{200m}$ and $r_{s}$ and found that scaling with $r_{200m}$ gives slightly better $\chi^{2}$ results than $r_{s}$. However, when the ratios of the simulation run $\chi^{2}$ over the mean ideal halo stacks $\overline{\chi^{2}}$ are taken (which will be further discussed in Section ~\ref{sec:stackData}), we get roughly similar results with both scaling methods. For consistency, throughout this paper we will scale with $r_{s}$.

We find that the resulting stacks using scaling even when using ideal NFW or DK haloes do not agree with the NFW or DK forms. The disagreement is due to parameter estimation of $M_{200c}$ and $c$, and therefore $r_{s}$.\footnote{This was tested by following this stacking process with one run using weak lensing parameter estimates and another run using true parameters} However this disagreement can be mitigated when using ratios of the $\chi^{2}$ results with respect to the $\overline{\chi^{2}}$ results of ideal DK haloes (see Sec.~\ref{sec:stackData}).

The errors of the stack at each $x$ can then be obtained from
\begin{equation}
\sigma_{\langle F \rangle}^{2}(x) = 
\frac{1}{2N^ {2} } \sum_{a=1}^{N_{c} }\sum_{s_{a};|\textbf{x}_{(a)s_{a} }| \in x'}^{}
\frac{w_{ (a, s_{a}) }^{2} \Sigma_{cr(a)}^{2} \epsilon_{(s_{a})}^{2}(\textbf{x}_{\rm s_{\rm a} })}
{4\rho_{c}^{2} ( M_{\rm (a)}, c_{\rm (a)}) r_{s}^{2}( M_{\rm (a)}, c_{\rm (a)})}.
\label{eq:sigmaForm}
\end{equation}
Both equations \ref{eq:stackForm} and \ref{eq:sigmaForm} are functions of Eq.~\ref{eq:stackx}.

\section{Results}
\label{sec:Results}


\subsection{Parameterized Mass Model Fits}
\label{sec:fits}

Throughout this paper we use the public module LmFit\footnote{The LmFit package is Free software, using an Open Source license} as the fitting tool \citep*{Newville2014}, where both the fits and the error bars are determined from the module. 
The cluster sample used in this section will be one of the LSST-like runs with true mass range for the sample of the 50 most massive clusters being $4.04 < M^{3D}_{200c} [10^{14} M_{\odot}]< 17.4$ with an average of $\langle M_{200}^{3D} [10^{14}M_{\odot}] \rangle = 6.64$. The true mass of a cluster, $M^{3D}$, is calculated by finding the particle deepest in the potential well of a cluster, and calculating the density within spherical shells around the particle. Once the average density becomes $200 \rho_{c}$, the corresponding radius is $r^{3D}_{200c}$ and the mass enclosed is $M^{3D}_{200c}$. There are no true concentrations calculated for the simulations for cluster-to-cluster comparisons.

\begin{figure}
	\includegraphics[width=\columnwidth]{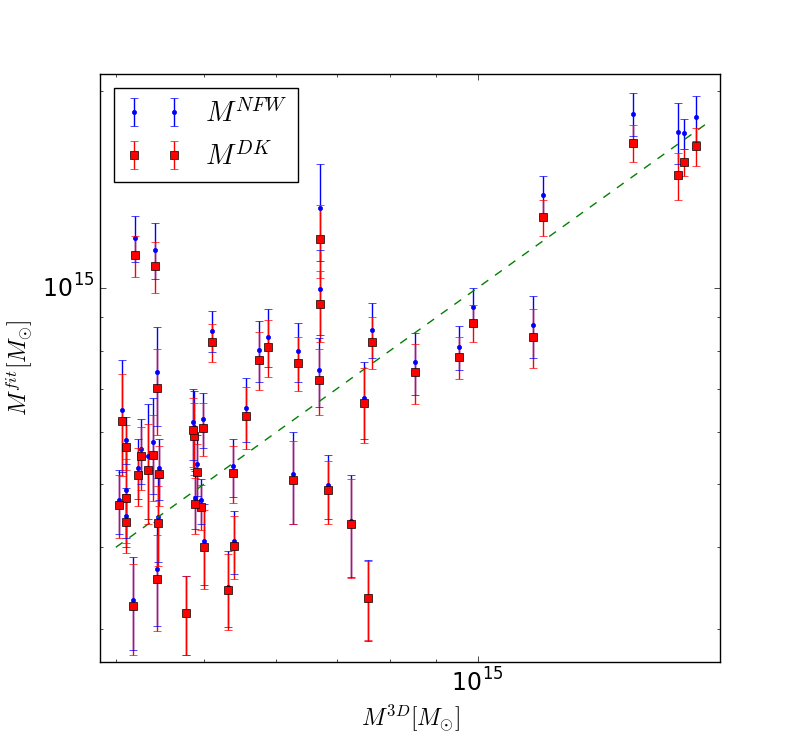}
	\caption{The red squares and blue points values plotted here are the DK and NFW mass estimates respectively versus the true masses from the simulation. The green dotted line is $y = x$. In this example, we consider the run with AGN 8.0 for LSST-like surveys and for the 50 most massive clusters in the sample. This shows that the DK fits prefer a lower mass estimate than that of the NFW. This trend is similar with other runs.
	}
	\label{fig:MfitsVsM3Ds}
\end{figure}

\begin{figure}
	\includegraphics[width=\columnwidth]{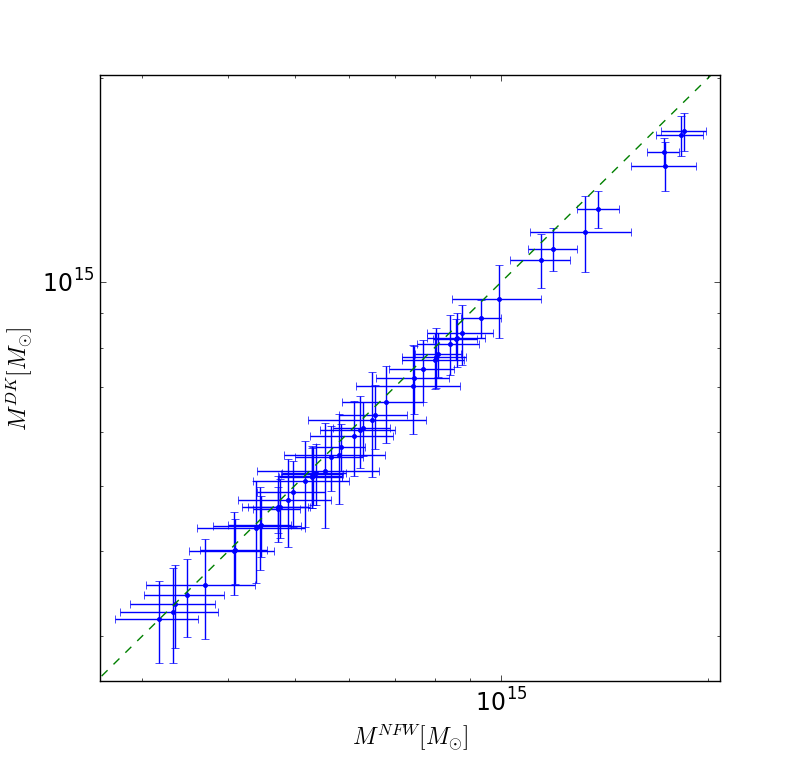}
	\caption{This plot is using the same data as Figure \ref{fig:MfitsVsM3Ds}. The green dashed line is $y = y$ and the blue points with error bars are the DK vs. NFW fit mass estimates. For the lower mass clusters (of the 50 most massive clusters in the simulation), the DK and NFW masses are in agreement. For higher masses, their estimates begin to diverge, showing that for the highest mass clusters DK mass estimates are lower than the NFW. 
	}
	\label{fig:MdksVsMnfws}
\end{figure}

Fig.~\ref{fig:MfitsVsM3Ds} and Fig.~\ref{fig:MdksVsMnfws} show individual mass estimates compared with true mass for an AGN 8.0 LSST-like run. The many other runs with different noise realizations yield similar results. The red squares and blue points and error bars are the DK and NFW $M_{200c}$ estimates respectively. The plots show that the DK and NFW fits agree well with one another for lower masses, and differ from one another for higher masses. For higher masses, the DK fits give lower masses than the NFW does. 

\begin{figure}
	\includegraphics[width=\columnwidth]{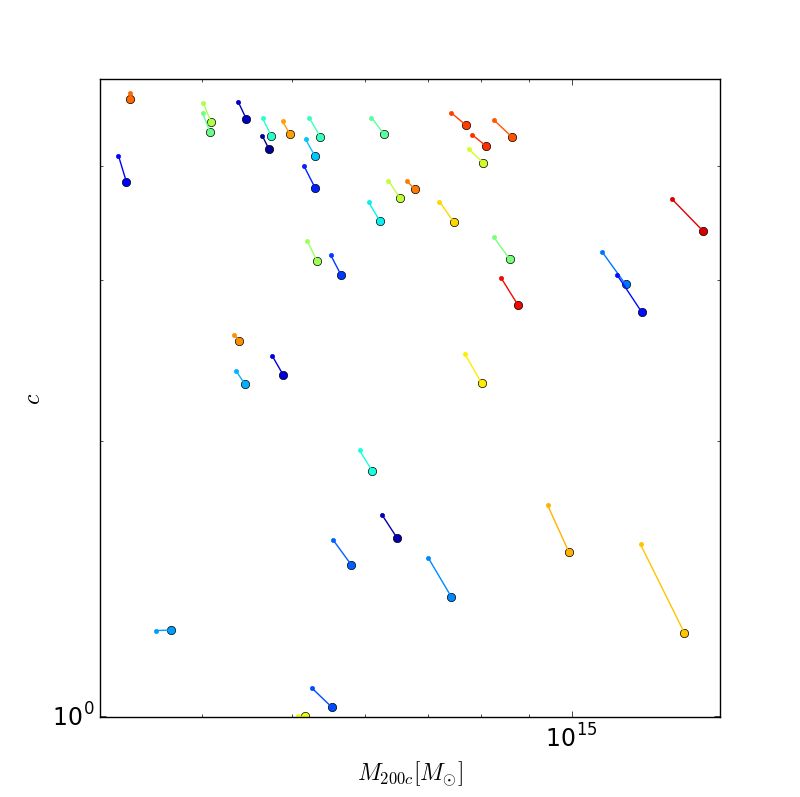}
	\caption{This figure is over the same run as Figures \ref{fig:MfitsVsM3Ds} and \ref{fig:MdksVsMnfws}.The NFW and DK fit parameters are marked by larger circles and smaller points respectively, where each cluster fit results are connected by lines of the same colour. Differences in fits in the x or y-directions reflects on the difference in estimates for mass and concentrations respectively. This plot shows that the NFW and DK profiles agree with mass and concentration estimation for lower mass bins (of the 50 most massive clusters) and the DK profile prefers a lower mass for higher mass bins, preferring higher concentrations to compensate. Also it looks like the DK profile generally prefers in a lower mass than the NFW throughout this cluster sample, though that is not always true in general. 
	}
	\label{fig:DKFits}
\end{figure}

Figure \ref{fig:DKFits} shows the NFW and DK fit parameters. The NFW and DK parameters are marked by circles and dots respectively, connected by lines. Differences in fits in the x or y-directions reflect on the difference in estimates for mass and concentrations respectively. 
For lower mass clusters the DK and NFW mass estimates are similar, which is what we would expect if the DK and NFW shear profiles are similar in this mass range (See Section \ref{sec:lensing}). For the largest mass clusters, the difference between the mass estimates tend to be higher. For the entire mass range we find that the DK mass estimates do better than the NFW. 
The overall bias (arithmetic average) of NFW mass estimates in this sample is $\approx 15\%$ (biased high), which is large in magnitude, but the sample includes two extremely triaxial clusters with the long axis close to the line-of-sight. Once these are excluded, the bias drops to $\approx 10\%$. For the DK the bias is $\approx 10\%$ (high bias) including all clusters and $\approx 5\%$ for excluding the two triaxial clusters. 
Throughout this paper we keep all of the 50 clusters in our stacking process and there is a significant scatter in the biases from run to run. When we have a larger sample of 300 clusters our bias drops significantly. For simulated clusters we found the bias for NFW and DK profiles to be overestimating by $\approx7\%$ and $\approx5\%$ respectively.

\subsection{Galaxy Cluster Triaxiality}
\label{sec:triaxiality}
\begin{figure}
	\includegraphics[width=\columnwidth]{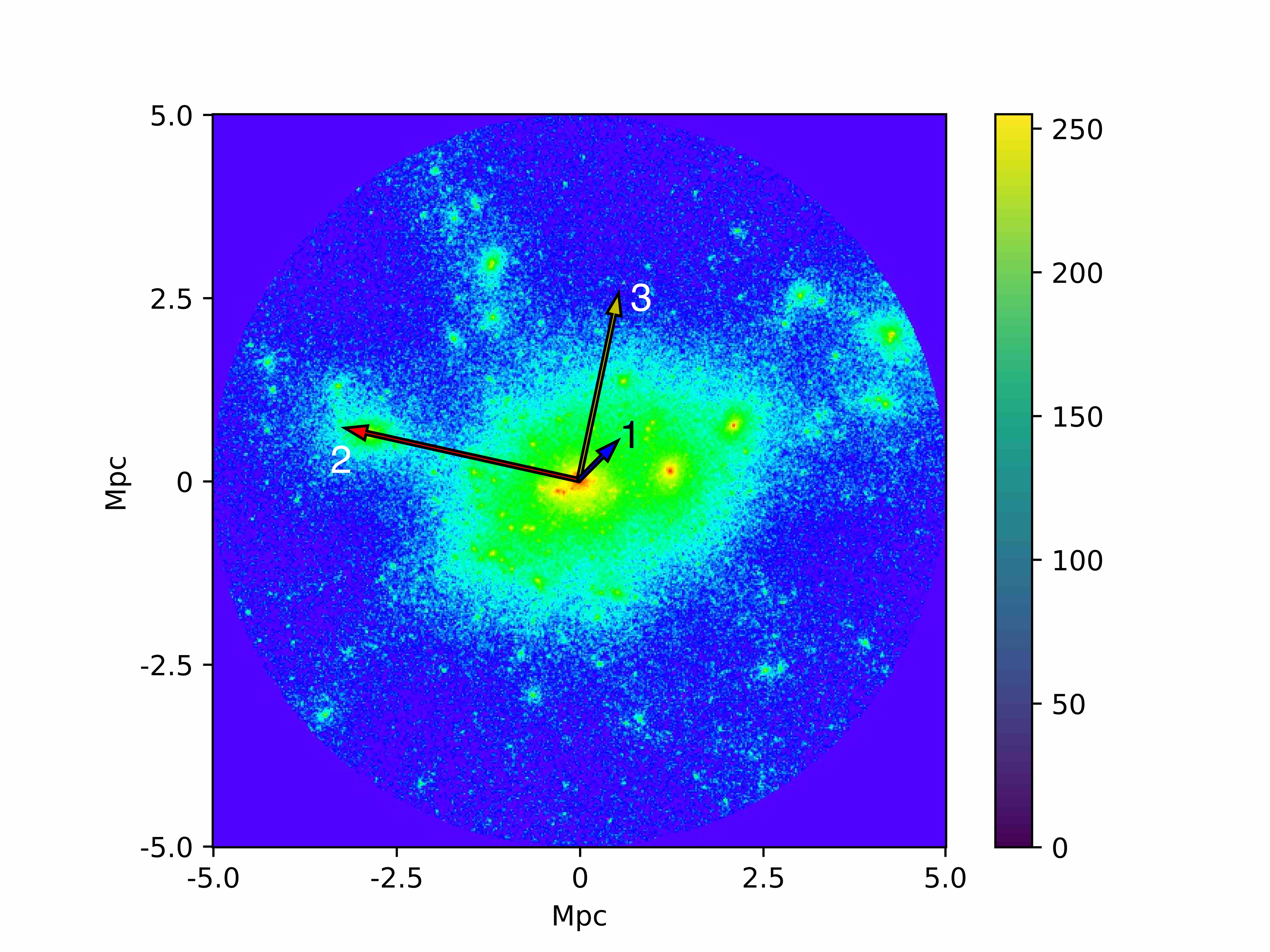}
	\caption{$\kappa$ map of a simulated cluster in xy projection overlaid with a vector representation of the
		cluster axes in projection. The $\kappa$ map is zoomed into a radius of 5 Mpc. The arrows labelled 1, 2, and 3 represent the major, intermediate, and minor axes respectively. The axes have been scaled
		according to their respective eigenvalues. Though longest in 3D space, the major axis appears shortest
		in projection because it lies in a similar direction to the line of sight (z axis).This leads to a large error in mass estimation when using a spherical
		model.
	}
	\label{fig:cluster15}
\end{figure}

Many of the clusters significantly depart from spherical, and as detailed in Lee et. al. (MNRAS submitted), the moments of inertia were calculated for each of the clusters, giving descriptions as ellipsoids with ratios for the major, intermediate and minor axes, along with their orientation in 3D and with respect to the z-axis along which the cluster mass is projected. Of the 50 most massive clusters we found that there are two clusters in this sample that are extremely triaxial, and their major axes are aligned close to the line of sight.  As noted above, when their synthetic lensing data are omitted, the average bias and error drop significantly. Figure \ref{fig:cluster15} shows a convergence map for one of the highly triaxial clusters. The three different arrows represent the physical projection of the major, intermediate, and minor axes in the xy-plane. Note that the shortest arrow corresponds to the major axis, indicating that the major axis is close to the line-of-sight as black, gray, and white respectively. From the moment of inertia tensor, the minor and intermediate to major axis ratios are $0.45$ and $0.62$ (Lee et. al (MNRAS submitted)). In addition the masses determined from the projections in the xz and yz-planes are significantly different. 
However, since in practice we would not be able to identify these clusters as highly triaxial, we keep them in our analysis. With huge samples of galaxy clusters expected from future surveys many clusters can be stacked to study mass profiles. The results from \cite{CorlessKing2009} indicate that stacking at least 100, and more ideally 500, clusters in a particular mass range would negate the impact of triaxiality on the determination of masses from a sample. 
We are currently investigating improved triaxial models for individual clusters, derived from lensing, X-Ray, Sunyaev-Zel'dovich and other cluster data.

\subsection{Ideal DK Lensing Data Sets}
\label{sec:idealDKHalos}
Before detailing the simulation results we explore how both stacking methods work with ideal DK haloes. We use the $50$ most massive clusters in the sample and determine their concentrations using DK14 $M-c$ relation. Then we generate ideal DK data sets and follow the steps in Section \ref{sec:sims} for LSST-like runs with realistic shape noise levels. We follow the steps in Sections \ref{sec:stackingWithoutScaling} and \ref{sec:stackingWithScaling}. 

\begin{figure}
	\includegraphics[width=\columnwidth]{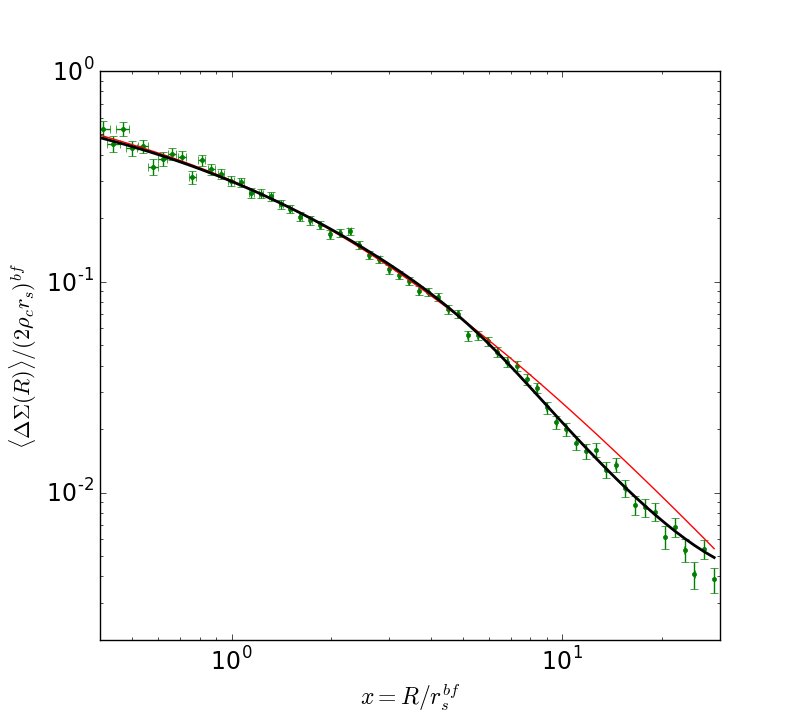}
	\caption{stack without scaling (Sections \ref{sec:stackingWithoutScaling}) for ideal DK haloes. The green points are  $\langle \Delta \Sigma(R) \rangle$ (stack without scaling) scaled by $2 \rho_{c} r_{s}$ (determined by fitting onto the stack), to compare with $F(x)$. The corresponding error bars is $\sigma_{\langle \epsilon \rangle}$. The thick black curve is $F^{DK}(x, \vec{\pi})$ (DK form) and, for reference, the red curve is $F^{NFW}(x)$ (NFW form).  
	This shows that stack without scaling for ideal DK haloes is well represented by the DK form.
	}
	\label{fig:idealDKWithoutScaling}
\end{figure}
In Figure \ref{fig:idealDKWithoutScaling} the green points are $\langle \Delta \Sigma(R) \rangle$ (stack without scaling), scaled by $2 \rho_{c} r_{s}$ determined by fitting onto the stack, with error bars $\sigma_{\langle \epsilon \rangle}$. The fit range for each stack depends on the type of run and can be found in Section \ref{sec:fitMethod}; in this case the fit range is set to $0.20 < r[$Mpc$] < 2.30$. The thick black curve is $F^{DK}(x, \vec{\pi})$ (DK form) and, for reference, the red curve is $F^{NFW}(x)$ (NFW form). The parameters used in $F^{DK}(x, \vec{\pi})$ is determined from the NFW fit results onto the stack ($r_{s} [$Mpc$] = 0.50$).
The stack uses 64 bins and the $\chi^{2}_{red}$ of the stack to $F^{DK}(x, \vec{\pi})$ is $1.02$. We use $d.o.f. = 64 + 2$ for the DK form because of the parameter, $\vec{\pi}$, dependency. This shows that $F^{DK}(x, \vec{\pi})$ represents the stacked signals of ideal DK haloes. This has been done over eight total runs with $\overline{\chi^{2}_{red} } = 1.00$.

\begin{figure}
	\includegraphics[width=\columnwidth]{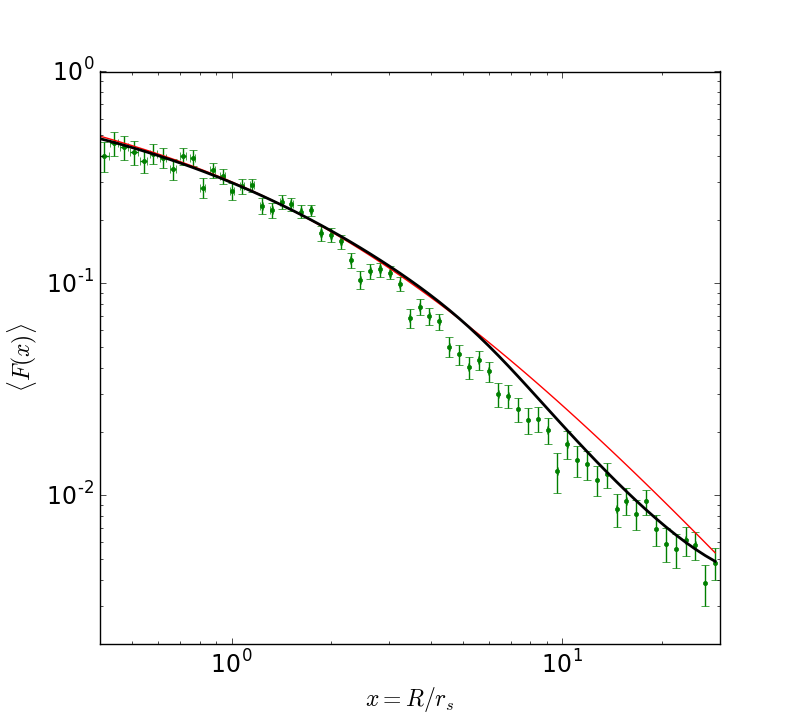}
	\caption{stack with scaling ($\langle F(x) \rangle$, Section \ref{sec:stackingWithScaling}) for ideal DK haloes. The green points are $\langle F(x) \rangle$ with the corresponding error bars $\sigma_{\langle F(x) \rangle}$. Even with ideal DK haloes, stack with scaling is not well represented by the DK form. This is likely due to the bias in parameter estimation that is needed for $\langle F(x) \rangle$ and parameter dependency in the DK form (i.e. each signal that goes into stack may have different concentrations and therefore different profile shapes, see Section \ref{sec:lensing}). The parameters used in the DK form is determined by fitting over the stack without scaling (Figure ~\ref{fig:idealDKWithoutScaling}).
	}
	\label{fig:idealDKWithScaling}
\end{figure}

In Figure \ref{fig:idealDKWithScaling} the green points are $\langle F(x) \rangle$ (stack with scaling) with error bars $\sigma_{\langle F \rangle}$. The thick black curve is $F^{DK}(x, \vec{\pi})$ (DK form) and, for reference, the red curve is $F^{NFW}(x)$ (NFW form). The parameters is the same as above (determined by NFW fitting the normal stack). $\chi^{2}_{red} = 4.10$ of the stack to $F^{DK}(x, \vec{\pi})$ with $d.o.f. = 64 + 2$, for the $64$ bins and the parameters $\vec{\pi}$ dependency. For eight total runs $\overline{\chi^{2}_{red}} = 8.11$. This shows that stack with scaling, using weak lensing mass and concentrations estimates, for ideal DK haloes is not well represented by $F^{DK}(x, \vec{\pi})$. 
 When we use the true parameters, the $\chi^{2}_{red}$ values are reduced, with an average of $1.63$.
 stack with scaling likely is not well represented by $F^{DK}(x, \vec{\pi})$ due to the difficulty in accurately estimating the parameters and that the shear form is dependent on parameters $\vec{\pi}$. The spread in parameters $\vec{\pi}$ values can be seen outside $r_{200c}$ ($=1.69 $ Mpc) (See Section \ref{sec:lensing}), where the stack dips below the NFW and DK forms. Since the shear form depends on $M_{200c}$ and $c$ outside $~r_{200c}$, the form varies for each individual signal outside of that. This may cause the stacked signals to differ from that of $F^{DK}(x, \vec{\pi})$ (DK form), where $c$ is determined from the fit onto $\langle \Delta \Sigma(R) \rangle$ (stack without scaling).

The DK form performs better for the stack without scaling, $\langle \Delta \Sigma(R) \rangle$ (then scaled by the best-fit parameters to the stack itself), as opposed to stack with scaling, $\langle F(x) \rangle$, for ideal DK haloes when using weak lensing parameter estimation. The DK form is better represented by stack with scaling when we use the exact parameters that created the haloes in the first place, but still not as good as stack without scaling. \textit{So for the rest of this paper we will focus on the stacking without scaling method, $\langle \Delta \Sigma(R) \rangle$. }

There is another DK scaled density form proposed by \cite{Umetsu2017} that gives would give good results when stacking very massive clusters. However, for the mass range considered here, we have tested that even though stacking with scaling works for ideal NFW haloes in the absence of noise, the individual cluster parameters are too poorly constrained with realistic noise for stacking to be viable.

\subsection{Stacking Without Scaling for AGN 8.0 Cluster Lensing Data}
\label{sec:stackData}

\begin{figure}
	\includegraphics[width=\columnwidth]{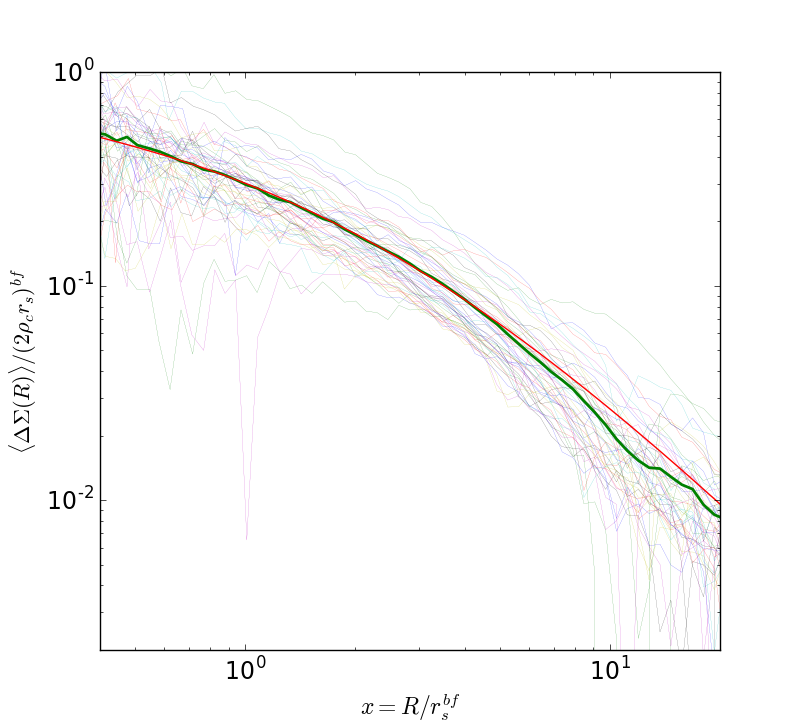}
	\caption{Each thin curve represents a cluster's reduced shear that went into the stack, the thick green curve represents the stack without scaling, and the thick red curve is the NFW form, $F^{\rm NFW}(x)$. Everything is then scaled using the best fit parameters onto the stack, to compare with $F(x)$. Here we follow Section \ref{sec:stackingWithoutScaling} with the 50 most massive clusters with no noise. For simplicity, we omitted error bars to show individual signals compared to the stack without scaling. This plot is an example of how stack without scaling compares with each individual cluster shear profile and the NFW form.
	}
	\label{fig:stackNNMM50WithoutScaling}
\end{figure}

In this section we look at the 50 most massive clusters in a single LSST-like run over the AGN 8.0 simulation while setting the radial bins as $0.20 < R' [$Mpc$] < 15.0$ in 64 bins. For illustration we look at the results with no shape noise, or $\epsilon^{s} = 0$ in Eq.~\ref{eq:ellipticity}.
In Fig.~\ref{fig:stackNNMM50WithoutScaling} we follow Sec.~\ref{sec:stackingWithoutScaling}. The thin curves are the individual cluster signals and the thick green curve is the stack without scaling ($\langle \Delta \Sigma(R) \rangle$, Eq.~\ref{eq:stackNiikura}), all scaled by the best fit parameters to the stack. The thick red curve is the same NFW form, $F^{\rm NFW}(x)$ in Equation \ref{eq:F}. Here, $\langle \Delta \Sigma (R) \rangle$ (stack without scaling) is scaled by $2 \rho_{c} r_{s}$ and $R$ is scaled by $r_{s}$, parameters determined by fitting to the stack (Equation \ref{eq:stackNiikura}), to compare with the DK form, $F^{DK}(x, \vec{\pi})$ (parameters $\vec{\pi}$ determined by the NFW fit onto the stack for $0.20 < r[$Mpc$] < 2.30$ in this case). Using the same parameters, each individual signal is scaled by $2 \rho_{c} r_{s} / \Sigma_{cr}$ and radial positions by $r_{s}$.

\begin{figure}
	\includegraphics[width=\columnwidth]{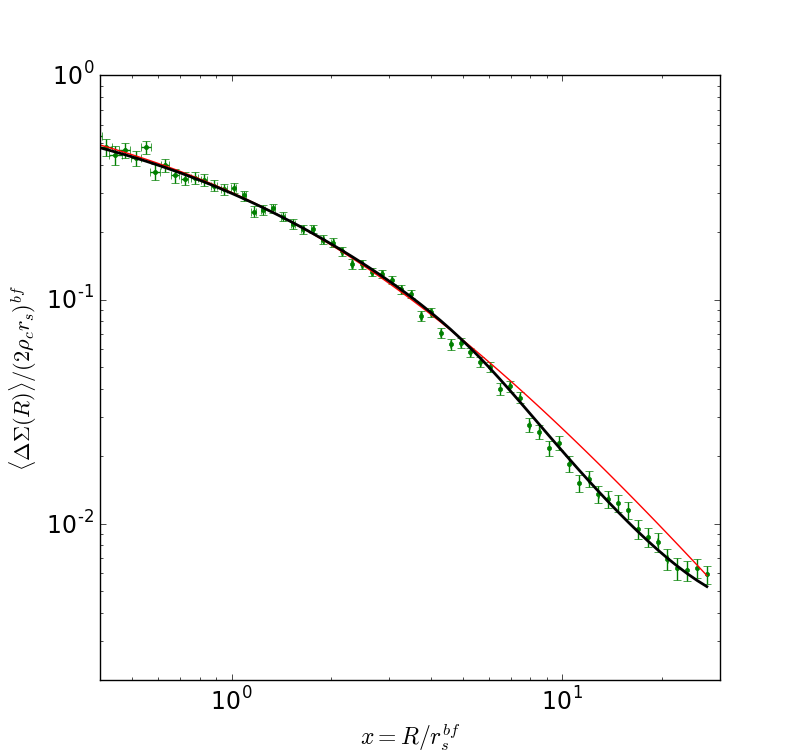}
	\caption{This plot uses realistic LSST-like clusters from the AGN 8.0 simulation. The green points are $\langle \Delta \Sigma(R) \rangle$ (stack without scaling, following Section \ref{sec:stackingWithoutScaling}), scaled by $(2 \rho_{c} r_{s})^{bf}$ (so we can compare with $F(x)$) with error bars of $\sigma_{\langle \epsilon \rangle} / (2 \rho_{c} r_{s})^{bf}$ (Eq.~\ref{eq:sigmaEllipticities}). The thick black curve is the DK form, $F^{DK}(x, \vec{\pi})$, and the red curve is the NFW form, $F^{NFW}(x)$. It is clear that the stack is better represented by DK than NFW. The DK $\chi^{2}_{red} = 1.01$ with $d.o.f. = 64 + 2$.
	}
	\label{fig:stackWOScaling}
\end{figure}

Now we include realistic shape noise ($\sigma_{\epsilon} = 0.25$ in Section \ref{sec:sims}) for the same clusters for one of the runs. These results can be seen in Figure ~\ref{fig:stackWOScaling}. This shows that the stack without scaling is very well represented by the DK profile, with $\chi^{2}_{red} = 1.01$ and $d.o.f. = 64 + 2$ (NFW: $\chi^{2}_{red} = 4.30$ and $d.o.f. = 64$).

\begin{table}
	\caption{Ratios for DK and NFW profile, $\chi^{2}_{red} / \overline{\chi^{2}_{red}}$, for the LSST-like runs for AGN 8.0 and DMO simulations. In this paper we only have $c$ as the free parameter for the DK form and we used $64$ radial bins for the analysis ($d.o.f. = 64 + 2$). The NFW form doesn't have parameters, so the $d.o.f. = 64$ for NFW. The average $\chi^{2}_{red}$ values for stack without scaling, $\langle \Delta \Sigma(R) \rangle$, for the ideal DK and NFW halo runs with LSST-like noise are $\overline{\chi^{2}_{red}} = 1.00$ and $1.26$ 
	 respectively. The $\overline{\chi^{2}_{red}}$ values are calculated over eight LSST-like runs but over ideal DK or NFW haloes.
}
	\label{symbols}
	\begin{tabular}{@{}lcccccc}
		\hline  & DK & & NFW & \\
		\hline
		Runs & AGN 8.0 & DMO & AGN 8.0 &  DMO\\
		\hline
		1 &  1.41 &  1.34 & 2.83 & 3.46\\ 
		2 &  1.01 &  1.20 & 3.41 & 2.50\\
		3 &  1.18 &  1.36 & 2.96 & 3.10\\
		4 &  1.20 &  1.35 & 2.00 & 3.47\\
		\hline
		Avg. & 1.20 & 1.31 & 2.84 & 3.13\\
		\hline
	
	\end{tabular}
	
	\medskip
		\label{table:mainStatistics}
\end{table}%

To help us determine if the stack is well represented by the DK profile, we use the NFW profile as reference. First, we follow the same procedure in determining the stacks but instead of using simulations we use ideal DK or NFW haloes. We then take an average of their $\chi^{2}_{red}$ values. To determine if the simulation haloes are more like DK or NFW, we take the ratio of the simulation stack $\chi^{2}_{red}$ values with the average of the ideal halo runs, $\overline{\chi^{2}_{red}}$.
Therefore, whichever ratio is closer to $1$ then the resulting simulation stack behaves more like either the ideal DK or NFW haloes.
In Table ~\ref{table:mainStatistics} we look at the DK and NFW results for AGN 8.0 and DMO simulations for LSST-like runs, without Large Scale Structure noise. With these ratios it is clear that the density profiles of the simulated clusters behave more like the DK profile, but using this analysis it is difficult to determine if baryonic prescriptions have any impact on their overall profile.

\subsection{Impact of Uncorrelated Large Scale Structure}

\begin{figure}
	\includegraphics[width=\columnwidth]{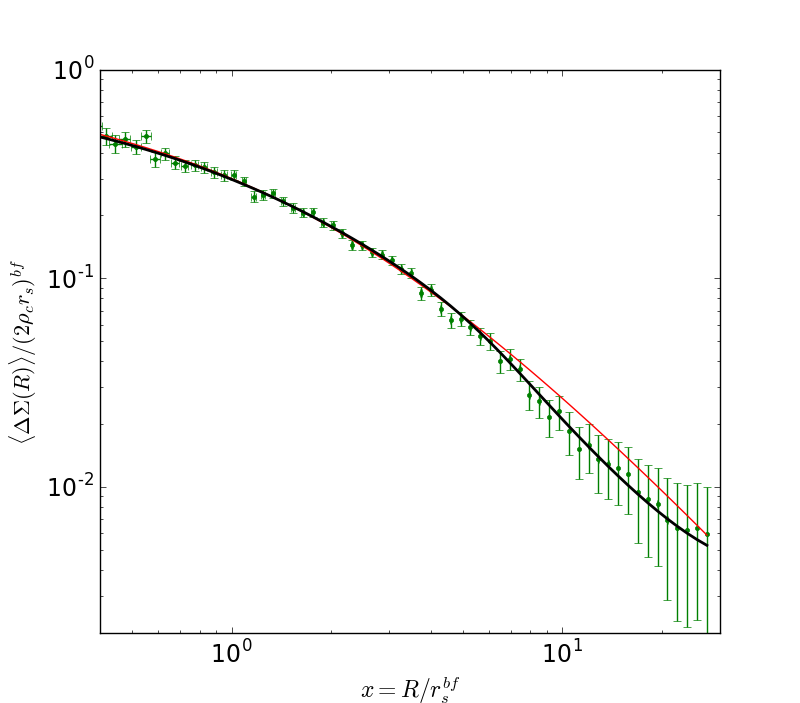}
	\caption{This plot uses realistic LSST-like clusters from the AGN 8.0 simulation but with added noise approximating that due to Large Scale Structure ($\sigma_{LSS}$). The green points are $\langle \Delta \Sigma(R) \rangle$ (stack with scaling, following Section ~\ref{sec:stackingWithoutScaling}), scaled by $(2 \rho_{c} r_{s})^{bf}$ (so we can compare with $F(x)$) with error bars as $\sigma_{\langle \epsilon \rangle} / (2 \rho_{c} r_{s})^{bf}$ (Eq.~\ref{eq:sigmaEllipticities}) and $\sigma_{LSS} / (2 \rho_{c} r_{s})^{bf}$ added in quadrature. The thick black curve is the DK form, $F^{DK}(x)$, and the red curve is the NFW form, $F^{NFW}(x)$. It is clear that the stack is better represented by DK than NFW. The added noise is extremely crude (flat $\sigma_{LSS} = 0.004$), but is included in this plot to show the difficulty in distinguishing between the DK and NFW forms when Large Scale Structure is considered. 
	}
	\label{fig:stackWOScaling_LSS}
\end{figure}
We also look at the results when Large Scale Structure noise $\sigma_{LSS}$ is included. 
In real data there would be more noise due to structure along the line of sight, which is not fully accounted for by taking $30$ Mpc boxes from the simulations. For simplicity we set $\sigma_{LSS} = 0.004$, typical of the noise at the outskirts of clusters due to LSS in \cite{Dodelson2004}; the inner region of the DK and NFW reduced shear forms are very similar anyway and the DK and NFW forms depart toward the outskirts. 
Another simplification in our treatment of the LSS is that the noise originating from uncorrelated projected LSS (i.e. structures not associated with the clusters themselves) is correlated at various scales, and this is of particular importance at large radii (in excess of $r \approx 10'$; e.g., \cite{Hoekstra2003}). 
Figure \ref{fig:stackWOScaling_LSS} is the results for the stacking without scaling method (Section \ref{sec:stackingWithoutScaling}), with new error bars with $\sigma_{LSS}$ added in quadrature to the stacked error bars, $\sigma^{2}_{\langle\epsilon\rangle}$. 
It is clear that the stack (without scaling) is well represented by the DK profile even when the noise due to large scale structure is included.
($\chi^{2}_{red} = 0.47$ and $0.84$ for DK and NFW forms respecively. For NFW, the $d.o.f. = 64$ while for DK $d.o.f. = 64 + 2$). The low $\chi^{2}_{red}$ values are due to large error bars.

\begin{table}
	\caption{Ratios for DK profile, $\chi^{2}_{red} / \overline{\chi^{2}_{red}}$, for the LSST-like runs for AGN 8.0 and DMO simulations. This table is similar to Table ~\ref{table:mainStatistics} but with Large Scale Structure noise $\sigma_{LSS} = 0.004$. The average $\chi^{2}_{red}$ values for stack without scaling, $\langle \Delta \Sigma(R) \rangle$, for the ideal DK and NFW halo runs with LSST-like noise are $\overline{\chi^{2}_{red}} = 0.54$ and $0.51$ 
	 respectively. The $\overline{\chi^{2}_{red}}$ values are calculated over eight LSST-like runs.}
	\label{symbols}
	\begin{tabular}{@{}lcccccc}
		\hline  & DK & & NFW & \\
		\hline
		Runs & AGN 8.0 & DMO & AGN 8.0 &  DMO\\
		\hline
		1 &  1.08 &  1.13 & 1.78 & 1.97\\ 
		2 &  0.87 &  0.96 & 1.64 & 1.49\\
		3 &  0.98 &  1.17 & 1.66 & 1.86\\
		4 &  0.92 &  0.97 & 1.35 & 1.66\\
		\hline
		Avg. & 0.97 & 1.06 & 1.61 & 1.75\\
		\hline
	
	\end{tabular}
	
	\medskip
		\label{table:mainStatistics_LSS}
\end{table}%

Table~\ref{table:mainStatistics_LSS} is similar to Table~\ref{table:mainStatistics}, but with the addition of noise due to Large Scale Structure. Again we can see that the simulation haloes are more like DK, but it is difficult to tell if baryonic prescriptions have any effect on the shape of them.

Overall the DK profile is a good representation of stacks created without scaling, $\langle \Delta \Sigma(R) \rangle$, of the most massive clusters in the simulations with LSST-like or WtG-like parameters. These conclusions for samples such as WtG will be strengthened with larger samples.

In this section we showed one example of a LSST-like run over the AGN 8.0 simulation with noise levels determined in Sections \ref{sec:stackingWithScaling} and \ref{sec:stackingWithoutScaling}. 
For this work, many variations of stacking with and without scaling, $\langle F(x) \rangle$ and $\langle \Delta \Sigma(R) \rangle$ respectively, have been tested against the DK and NFW forms, $F^{DK}(x, \vec{\pi})$ and $F^{NFW}(x)$ respectively. We see how the inclusion of ellipticity amplitude in the weight (Equation \ref{eq:weight}, as in \cite{Niikura2015}) and how using ``true parameters" (or better estimated parameters) as opposed to weak lensing fit parameters effects the overall results. Using ellipticity amplitude in the weight, as in \cite{Niikura2015}, has little effect on the $\chi^{2}$ ratios. Using the ``true parameters" does give better $\chi^{2}_{red}$ values but the ratios and plots are qualitatively the same. We also explore increasing $\sigma_{\epsilon}$, from $0$ to $0.25$, for the Gaussian distribution where we randomly choose the intrinsic ellipticities (See Section \ref{sec:sims}) to see how it effects stack with scaling, $\langle F(x) \rangle$. We find that stack with scaling results still aren't represented by the DK form due to the parameters $\vec{\pi}$ dependency in $F^{DK}(x, \vec{\pi})$.

\subsection{Application of Stacking Without Scaling to Extended Field-of-View WtG-like Data}
\label{sec:WtG}
\begin{figure}
	\includegraphics[width=\columnwidth]{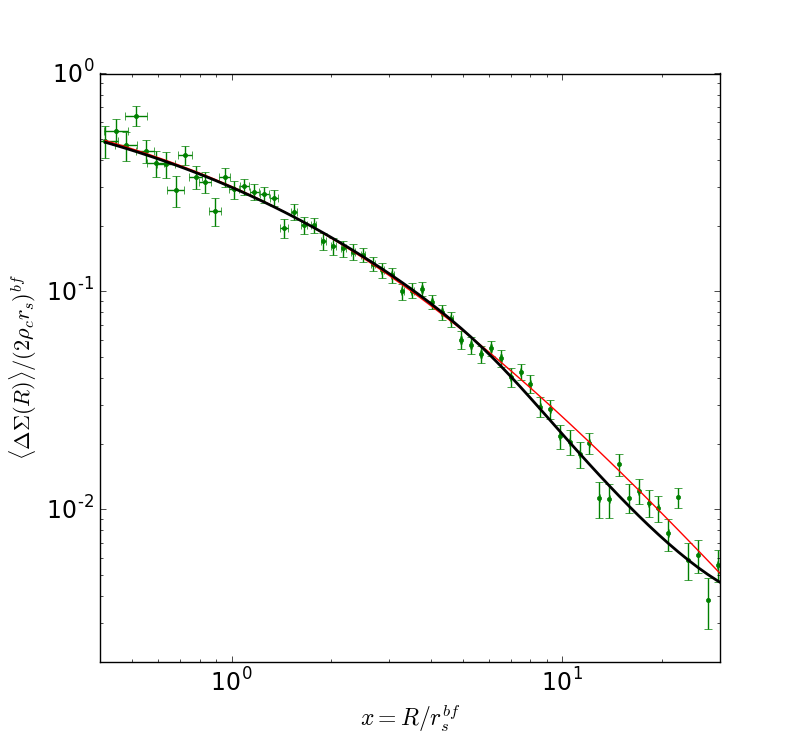}
	\caption{Here we plot stack without scaling ($\langle \Delta \Sigma(R) \rangle$, following Section ~\ref{sec:stackingWithoutScaling}) in green for a WtG-like run for the AGN 8.0 simulation using the 50 most massive clusters. The thick black and red curves are $F^{DK}(x, \vec{\pi})$ and $F^{NFW}(x)$, the DK and NFW forms, respectively. This shows that with a smaller number density of background galaxies, $n_{0}$, the more difficult it is to distinguish between the NFW and DK forms (Compare with Figure \ref{fig:stackWOScaling}, where the number density is $n_{0} = 30$ gals/arcmin$^{2}$). So in order to distinguish, we either need to increase the number of sources or include more cluster signals.
	}
	\label{fig:stackWtG}
\end{figure}

For WtG-like runs we use $n_{0} = 10$ gals/arcmin$^{2}$ instead of LSST-like runs, where $n_{0} = 30$ gals/arcmin$^{2}$. The field-of-view used in our simulation is $30$ Mpc, more than double that of typical clusters from the WtG survey.
Figure~\ref{fig:stackWtG} shows stack without scaling ($\langle \Delta \Sigma(R) \rangle$) in green for a WtG-like run for the AGN 8.0 simulation using the 50 most massive clusters, compared to the DK and NFW forms, the thick black and red lines respectively. To show how the number density effects error bars, we exclude large scale structure noise. 
With a decrease in number of background sources, the error bars increase (compare with Figure \ref{fig:stackWOScaling}, where the number density is $n_{0} = 30$ gals/arcmin$^{2}$), rendering it the NFW and DK forms indistinguishable from the stack. In order to distinguish between the NFW and DK profiles, we need to either increase the number density of background galaxies or include more clusters.

\begin{table}
	\caption{Ratios for DK and NFW profile, $\chi^{2}_{red} / \overline{\chi^{2}_{red}}$, for the WtG-like runs for AGN 8.0 and DMO simulations. The DK form has $\vec{\pi}$ as the free parameters and uses $64$ radial bins for the analysis, so the $d.o.f. = 64 + 2$. The NFW form does not have any parameters in the form, so the $d.o.f. = 64$. The average $\chi^{2}_{red}$ values for stack without scaling, $\langle \Delta \Sigma(R) \rangle$, for the ideal DK and NFW halo runs with WtG-like noise are $\overline{\chi^{2}_{red}} = 0.98$ and $1.13$ 
	 respectively. The $\overline{\chi^{2}_{red}}$ values are calculated over eight WtG-like runs but over ideal DK or NFW haloes.}
	\label{symbols}
	\begin{tabular}{@{}lcccccc}
		\hline  & DK & & NFW & \\
		\hline
		Runs & AGN 8.0 & DMO & AGN 8.0 &  DMO\\
		\hline
		1 &  1.33 &  1.35 & 1.14 & 2.12\\ 
		2 &  1.37 &  0.95 & 1.21 & 1.72\\
		3 &  1.05 &  1.26 & 2.18 & 1.54\\
		4 &  0.73 &  1.21 & 1.47 & 1.28\\
		\hline
		Avg. & 1.12 & 1.19 & 1.50 & 1.67\\
		\hline
	
	\end{tabular}
	
	\medskip
	
	\label{table:WtGStatistics}
\end{table}%

\begin{table}
	\caption{Ratios for DK profile, $\chi^{2}_{red} / \overline{\chi^{2}_{red}}$, for the WtG-like runs for AGN 8.0 and DMO simulations. This table is similar to Table \ref{table:WtGStatistics} but with Large Scale Structure noise $\sigma_{LSS} = 0.004$. The average $\chi^{2}_{red}$ values for $\langle \Delta \Sigma(R) \rangle$ for the ideal DK and NFW halo runs with WtG-like noise are $\overline{\chi^{2}_{red}} = 0.62$ and $0.65$ 
	 respectively. The $\overline{\chi^{2}_{red}}$ values are calculated over eight WtG-like runs.}
	\label{symbols}
	\begin{tabular}{@{}lcccccc}
		\hline  & DK & & NFW & \\
		\hline
		Runs & AGN 8.0 & DMO & AGN 8.0 &  DMO\\
		\hline
		1 &  1.34 &  1.59 & 1.35 & 2.07\\ 
		2 &  1.12 &  0.91 & 1.11 & 1.29\\
		3 &  0.96 &  1.05 & 1.48 & 1.24\\
		4 &  0.76 &  0.99 & 1.14 & 1.08\\
		\hline
		Avg. & 1.05 & 1.14 & 1.27 & 1.42\\
		\hline
	
	\end{tabular}
	
	\medskip
		\label{table:WtGStatistics_LSS}
\end{table}%

Tables \ref{table:WtGStatistics} and \ref{table:WtGStatistics_LSS} are similar to the LSST-like runs at the end of Section~\ref{sec:stackData}.
Overall the results are that the DK profile performs better than the NFW when representing the simulation stacks, though the differences between the ratios are somewhat smaller. So even though the smaller number density of sources and the large scale structure noise both make the $\chi^{2}_{red}$ values larger, the ratios ($\chi^{2}_{red} / \overline{\chi^{2}_{red}} \approx 1$) can be used to show that the WtG-like simulations behave more closely to that of DK than NFW haloes.

\begin{figure}
	\includegraphics[width=\columnwidth]{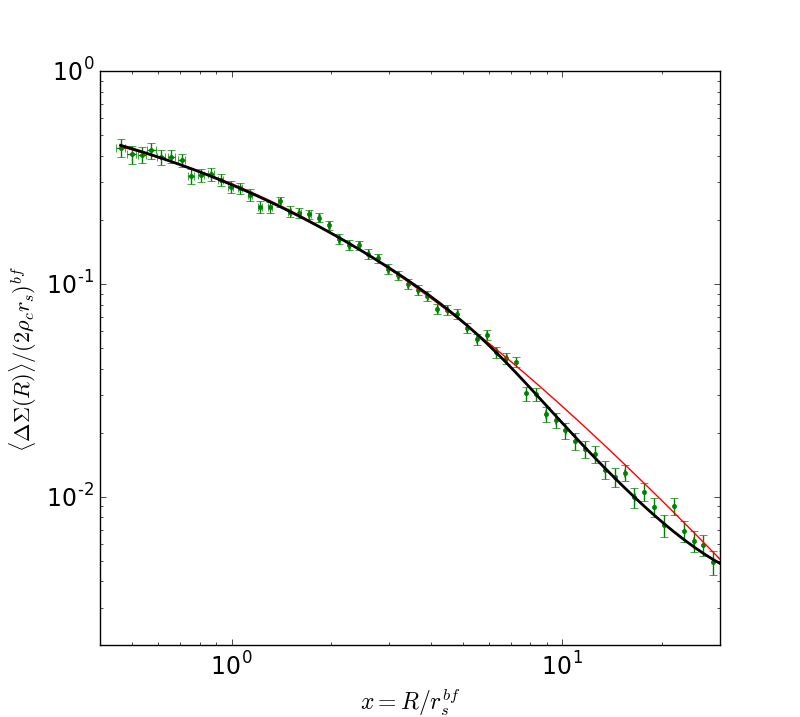}
	\caption{This figure is the same as Figure \ref{fig:stackWtG} but with $300$ (as opposed to $50$) of the most massive clusters from the AGN 8.0 simulation for a WtG-like run ($n_{0} = 10$ gals/arcmin$^{2}$). $\langle M_{200c}^{3D} \rangle = 3.30 \times 10^{14} \mathrm{M_{\odot}}$ with $1.97 < M_{200c}^{3D} [10^{14} \mathrm{M_{\odot}}] < 17.35$. This figure shows the results for a larger number of clusters with a much larger mass range, where $\chi^{2}_{red}$ are $0.91$ and $1.33$ for DK ($d.o.f. = 64 + 2$) and NFW ($d.o.f. = 64$) respectively.
	}
	\label{fig:stackWtG300}
\end{figure}

Another way that the $\chi^{2}_{red}$ values can be improved is to simply increase the number of clusters that go into the stack. Throughout this paper we have used the 50 most massive clusters from the simulations. Here we show the results of the 300 most massive $M_{200c}^{3D}$ values in the AGN 8.0 simulation for a WtG-like run ($n_{0} = 10$ gals/arcmin$^{2}$). $\langle M_{200c}^{3D} \rangle = 3.30 \times 10^{14} \mathrm{M_{\odot}}$ with $1.97 < M_{200c}^{3D} [10^{14} \mathrm{M_{\odot}}] < 17.35$. 
Here we exclude the noise due to large scale structure to show how the increase of number of clusters in the stack can effect the error bars. 
Figure \ref{fig:stackWtG300} shows that, even with the large mass range and lower number density of sources, stacked shear can be used to show that haloes behave more like DK than NFW haloes on large scales, so long as there is a sufficient number of cluster signals to make up for the lower number density of sources (Compare with Figure \ref{fig:stackWtG}, where there are 50 clusters in the stack). 
In this case we just use the $\chi^{2}_{red}$ values; $\chi^{2}_{red}$ are $0.91$ and $1.33$ for DK ($d.o.f. = 64 + 2$) and NFW ($d.o.f. = 64$) respectively. 

Since there are fewer galaxy clusters in the CLASH sample than in the WtG sample, we concur with the conclusions \cite{Umetsu2017} regarding the CLASH sample. Even with a wider field of view for a sample such as WtG, it is unlikely that the large scale environment of the clusters could be studied at a level to distinguish between the DK and NFW profile.

\section{Conclusions}
\label{sec:conclusions}
Gravitational lensing is an essential tool for probing the distribution of mass in galaxy clusters, most of which is dark matter. The most commonly used mass density profile to describe cluster scale structures is the NFW profile, which has stood two decades of scrutiny. Advances in the resolution of cosmological simulations, and in the incorporation of physical processes associated with baryonic matter - beyond gravity for dark matter only simulations, have resulted in refinements to the NFW model, for example that of \cite{DK2014}. 

Gravitational lensing estimates of cluster mass have rather a large scatter, with sources of noise including the finite sampling of the lensing potential by background galaxies which have an intrinsic distribution of shapes. Other factors include triaxiality in 3D, particularly when clusters are very elongated along the line of sight or in the plane of the sky. Besides boosting the lensing signal, and hence the quality of information on the average cluster mass profile, stacking averages the 3D structure of clusters, assuming that the sample is random on the sky.

In this paper we used 50 clusters extracted the cosmological simulations from cosmo-OWLS \cite{LeBrun2014}, specifically DMO and AGN 8.0 runs, the latter implementing feedback from black holes and other baryonic physics. We considered synthetic weak lensing data with background galaxy number density characteristic of the Weighing The Giants survey, and future LSST surveys. Using two different stacking procedures, we compared the accuracy with which NFW and DK models describe the stacked lensing data, and the prospects for measuring a departure from the NFW form. On larger scales, and for stacked lensing data, the DK model gives a more accurate description of the azimuthally averaged shear from WtG-like and LSST-like surveys. 
In particular for the LSST-like surveys, assuming $n_{0} = 30$ gals/arcmin$^{2}$ compared with the $n_{0} = 10$ gals/arcmin$^{2}$ for WtG, there are good prospects for detecting features beyond the applicability of the NFW model, in particular the steepening of the density profile around the splashback radius. The conclusions are the same for the DMO and AGN 8.0 runs. This is consistent with detailed studies of the lensing signatures of individual clusters from these and other cosmo-OWLS runs (Lee et al. (MNRAS submitted)).

The distinction between the lensing signals of NFW and DK models has interesting implications for the estimation of cluster mass, and constraints on the dark matter and other large scale structure surrounding them. As discussed in \cite{DK2014}, the parameters of their model are sensitive to the rate at which matter is accreted onto clusters. With large enough samples of clusters, we may be able to identify subsets of those which are more or less rapidly accreting, based on some physical indicator such as blue stellar populations indicative of star formation, and test whether they have distinct DK parameters and the relationship between splashback radius and mass accretion rate. In future work we will stack synthetic and real lensing data sets aligned based on their longest axes on the sky, since filaments tend to preferentially occur close to the major axes of clusters. Preliminary simulations indicate that this will enable us to study the periphery of clusters, and the large scale structures in which they are embedded, in greater detail.

\section*{Acknowledgements}

MWF, BEL, and LJK acknowledge support for this work by National Aeronautics and Space Administration Grant No. NNX16AF53G.

RLB acknowledges support from the NSF REU program at the Maria Mitchell Observatory.

DA recognizes support by the Kavli Institute for Cosmological Physics at the University of Chicago through grant NSF PHY-1125897 and an endowment from the Kavli Foundation and its founder Fred Kavli.

We would like to thank Regina Jorgenson for advising RLB at MMO's REU program and for helpful discussions.




\bibliographystyle{mnras}
\bibliography{References}

\appendix


\bsp	
\label{lastpage}
\end{document}